\documentclass[a4paper,11pt]{article}
\pdfoutput=1 

\usepackage{jheppub} 

\usepackage[OT1]{fontenc} 
\graphicspath{{figs/}} 	     
\usepackage{booktabs} 
\usepackage{bbold} 
\usepackage{mathtools}
\usepackage{multirow}
\usepackage[version=4]{mhchem} 
\usepackage{physics}
\usepackage{pdflscape}
\usepackage{slashed}
\usepackage{xcolor}
\usepackage{placeins} 
\usepackage{tikz} 
\usepackage{tikz-feynman} 
\usepackage{tikz-feynhand} 
\tikzfeynmanset{compat=1.1.0} 

\usepackage{feynmp-auto}
\usepackage{graphicx}

\usepackage{ulem}

\newcommand{\bea} {\begin{eqnarray}}
\newcommand{\eea} {\end{eqnarray}}

\DeclareMathOperator{\GeV}{\text{GeV}}

\title{Heavy Neutral Lepton at Same-Sign Muon Collider}

\author[a]{Ryuichiro Kitano,}
\author[b,c]{Ian Low,}
\author[a]{Ryutaro Matsudo,} 
\author[a,d,e]{Shohei Okawa,} 
\author[c]{and Subhojit Roy}

\affiliation[a]{Yukawa Institute for Theoretical Physics, Kyoto University, Kyoto 606-8502, Japan}
\affiliation[b]{Department of Physics \& Astronomy, Northwestern University, Evanston, IL 60208, USA}
\affiliation[c]{High Energy Physics Division, Argonne National Laboratory, Lemont, IL 60439, USA}
\affiliation[d]{Asia Pacific Center for Theoretical Physics, Pohang, 37673, Korea}
\affiliation[e]{Department of Physics, Pohang University of Science and Technology, Pohang, 37673, Korea\\}

\emailAdd{ryuichiro.kitano@yukawa.kyoto-u.ac.jp}
\emailAdd{ilow@northwestern.edu}
\emailAdd{matsudo@post.kek.jp}
\emailAdd{shohei.okawa@apctp.org}
\emailAdd{sroy@anl.gov}

\abstract{We explore the discovery potential of heavy neutral leptons (HNLs), motivated by models addressing the origin of neutrino masses, at the proposed high-energy same-sign muon collider known as $\mu$TRISTAN. The study focuses on two complementary HNL-mediated signatures:  (i) the lepton-flavor-violating (LFV) channel $\mu^+\mu^+ \to W^+\tau^+\bar\nu_\mu$ and (ii) the lepton-number-violating (LNV) channel $\mu^{+}\mu^{+} \to W^{+}W^{+}$. The LNV process is the muon analogue of inverse neutrinoless double beta decay and, if observed, would provide strong evidence for Majorana neutrinos, while the LFV process offers a novel probe of flavor-changing neutral currents in the lepton sector. At the $\mu$TRISTAN collider with $\sqrt{s} \sim \mathcal{O}(10)~\text{TeV}$,
the resulting sensitivity to the HNL mixing with muon and tau neutrinos, as a function of mass, can surpass current bounds from the measurements of electroweak precision observables over a broad mass range. In particular, for the mixing with muon neutrinos, the collider bound can improve by an order of magnitude for $5$--$10$~TeV HNLs.}

\begin{document}
\begin{flushright}
     YITP-25-166
\end{flushright}
\maketitle
\flushbottom 
\section{Introduction}
The discovery of the Higgs boson at a mass of approximately 125 GeV by the ATLAS and CMS collaborations at the Large Hadron Collider (LHC)~\cite{ATLAS:2012yve, CMS:2012qbp} marked a major milestone in particle physics, confirming the mechanism of electroweak symmetry breaking and completing the particle content of the Standard Model (SM).
As the LHC enters its high-luminosity phase, no clear evidence of physics beyond the SM has yet been observed. A key focus of the High-Luminosity LHC (HL-LHC) will be to continue the search for new physics and to investigate the properties of the Higgs boson in greater detail through precision measurements of its couplings.
Despite the SM’s remarkable success, it remains incomplete. It fails to account for several fundamental phenomena, including the nature of dark matter, the origin of neutrino masses, the observed matter–antimatter asymmetry in the universe, and the flavor structure of fermions. The lack of new physics signals so far has motivated the exploration of complementary experimental strategies and next-generation collider facilities aimed at addressing these open questions.

In light of this, a wide range of proposed future colliders aims to extend the reach of the LHC. Lepton colliders, in particular, offer clean experimental environments for precision studies of the Higgs sector and electroweak interactions, while also providing complementary sensitivity to beyond-the-SM (BSM) scenarios. Facilities under consideration include the Future Circular
Collider (FCC)~\cite{FCC:2025lpp}, the International Linear Collider (ILC)~\cite{Behnke:2013xla,Adolphsen:2013jya,Adolphsen:2013kya,Bambade:2019fyw,Zarnecki:2020ics}, the Linear Collider Facility (LCF)~\cite{LinearCollider:2025lya}, the Compact Linear Collider (CLIC)~\cite{Aicheler:2018arh,Zarnecki:2020ics}, the Cool Copper Collider (C$^3$)~\cite{Vernieri:2022fae,Breidenbach:2023nxd}, the HALHF project~\cite{Lindstrom:2023owp,Foster:2023bmq}, the Circular Electron Positron Collider (CEPC)~\cite{CEPCStudyGroup:2018ghi}, LEP3~\cite{Anastopoulos:2025jyh} and the Muon Collider~\cite{Accettura:2023ked}.

Among these proposals, muon colliders have emerged as a very interesting future
possibility. They combine the advantages of point-like collisions with the ability to reach multi-TeV center-of-mass energies, while avoiding significant synchrotron radiation losses~\cite{Heusch:1995yw,Antonelli:2015nla,Long:2020wfp,Delahaye:2019omf,Accettura:2023ked,MuonCollider:2022nsa, Begel:2025ldu}. Traditionally, muon colliders involve opposite-sign beams ($\mu^{+} \mu^{-}$), but recent advances in ultra-cold muon technology, particularly from the J-PARC facility~\cite{Kondo:2018rzx, Aritome:2024rlu}, have enabled the possibility of constructing high-intensity same-sign muon ($\mu^{+} \mu^{+}$) beams. This forms the basis of the $\mu$TRISTAN concept, a proposed multi-TeV same-sign muon collider that utilizes ultra-cold anti-muons. 
Such a machine could reach sufficiently high luminosity and energy to support a wide range of precision and discovery physics.
Although the $\mu^+ \mu^+$ collider lacks the ability
of producing new particles
via $s$-channel annihilations,
it provides a unique opportunity
to explore the lepton sector
by having a non-vanishing 
lepton number in the initial state.
Unlike conventional colliders, it can also search for exotic interactions that are difficult to access elsewhere. Beyond its role as a Higgs factory, $\mu$TRISTAN offers exciting opportunities to investigate a broad range of scenarios beyond the Standard Model~\cite{Belanger:1995nh,Gluza:1995ix,Fridell:2023gjx,Fukuda:2023yui,Das:2024gfg, Kriewald:2024cnt, deLima:2024ohf, Hamada:2022mua, Das:2024kyk}.
Thus, 
if the $\mu^+$ cooling facility
can be built somewhat earlier, there is certainly an option
to start with the $\mu^+ \mu^+$
collider where many physics programs, including the Higgs precision measurements, are
possible,
and one can wait for the $\mu^-$ cooling
to come later to explore
the energy frontier.

The observation of tiny but nonzero neutrino masses and flavor mixing~\cite{ParticleDataGroup:2020ssz} presents one of the most compelling signs of BSM physics. Among the simplest and most well-motivated extensions of the SM that address this puzzle are models introducing Heavy Neutral Leptons (HNLs), the SM-singlet fermions that couple weakly to the active neutrinos and participate in the generation of neutrino masses through seesaw mechanisms. In particular, the type-I seesaw~\cite{Minkowski:1977sc,Yanagida:1979as,Glashow:1979nm,Gell-Mann:1979vob,Mohapatra:1979ia}, and its variants such as the linear~\cite{Akhmedov:1995vm, Barr:2003nn,Malinsky:2005bi} and inverse seesaw~\cite{Schechter:1980gr, Gronau:1984ct, Mohapatra:1986bd}, provide elegant frameworks in which small neutrino masses arise naturally from the suppression of lepton-number-violating (LNV) mass scales.
These sterile states also appear in broader new physics constructions, including left-right symmetric models, radiative neutrino mass models, and Grand Unified Theories, and can span a wide range of masses depending on their theoretical origin~\cite{Deppisch:2015qwa,Bolton:2019pcu}. In addition to their role in neutrino mass generation, HNLs have been studied as candidates for baryogenesis via leptogenesis, and as possible contributors (partly) to the dark matter relic abundance. The LHC has actively pursued searches for such states, particularly in final states involving same-sign dileptons or multileptons with missing energy~\cite{CMS:2018iaf,CMS:2022hvh,CMS:2022fut}, placing bounds on their masses and mixings. However, the hadronic environment of the LHC limits the sensitivity to rare LNV processes, motivating the need for cleaner experimental setups.
Future lepton colliders and in particular, the proposed $\mu$TRISTAN $\mu^+ \mu^+$  collider, offer a promising platform for probing HNLs with enhanced sensitivity. The unique $\mu^+ \mu^+$  initial state suppresses many SM backgrounds, improving discovery prospects for Majorana neutrinos and other BSM scenarios involving extended neutrino sectors. Given the central role that HNLs play in connecting neutrino physics, cosmology, and high-energy theory, dedicated searches at future facilities like $\mu$TRISTAN are well-motivated and timely.

One particularly intriguing process discussed in the context of $\mu^+ \mu^+$ colliders is the LNV reaction $\mu^+ \mu^+ \to W^+ W^+$ \cite{Heusch:1995yw}. This process can be viewed as the muon analogue of ``inverse neutrinoless double beta decay"~\cite{Rizzo:1982kn}, providing a direct probe of LNV in the muon sector.\footnote{The analogy to inverse neutrinoless double beta decay arises because both processes involve two same-sign leptons annihilating into two same-sign $W$ bosons via an intermediate Majorana neutrino, violating lepton number by two units. While conventional $0\nu\beta\beta$ experiments target LNV in the electron sector through nuclear decays, they are not sensitive to LNV involving other lepton flavors. In contrast, same-sign lepton colliders offer a clean environment to probe flavor-specific LNV processes, such as $\mu^+ \mu^+ \to W^+ W^+$, and can explore scenarios where new physics couples more strongly to muons than to electrons~\cite{Tornow:2014vta,EXO-200:2019rkq,GERDA:2020xhi,CUORE:2021mvw,KamLAND-Zen:2022tow, Rodejohann:2010jh}. Thus collider-based searches provide a complementary avenue to nuclear-based experiments, particularly in regions of parameter space involving heavy mediators beyond the kinematic reach of double beta decay.} Thus, if this process is observed,
this would be a strong indication
that the neutrinos are
the Majorana particle.
Given that the muon collider
can go up to ${\cal O}(10)$~TeV center
of mass energy
with a ${\cal O}(10)$~km level collider ring,
the search for this process
can reach unexplored regions
of the parameter space
in various neutrino mass models~(see related works in Refs.~\cite{Jiang:2023mte, Das:2024kyk, Yang:2023ojm,deLima:2024ohf,Dehghani:2025xkd, Fridell:2023gjx, Dev:2023nha, Han:2026ufa}).

In this work, we study the model of the HNLs
at the $\mu^+ \mu^+$ collider. We consider two processes: $\mu^+\mu^+ \to W^+\tau^+\bar\nu_\mu$ and $\mu^+\mu^+\to W^+W^+$.
The first process is independent of the Majorana mass term and preserves the lepton number symmetry while violating the lepton flavor symmetry.
The second process is analogous to inverse neutrinoless double beta decay at $e^-e^+$ colliders, violating the lepton number symmetry.
We find that both processes provide stronger bounds than the existing limits from electroweak precision measurements on the corresponding parameters, for a center-of-mass energy of 10 TeV, and an integrated luminosity of $1~\mathrm{ab}^{-1}$.

The paper is organised as follows: In Sec.~\ref {sec:model}, we outline the general setup of the HNL framework and review the existing experimental constraints, particularly those from the measurements of the electroweak precision observables (EWPOs). In Sec.~\ref{HNLsearch}, we describe the search strategies for HNLs at a same-sign muon collider, focusing on both lepton-flavor–violating (LFV) and LNV processes. We present the details of our simulation setup, event selection, and sensitivity projections. Our main results are summarised and discussed in Sec.~\ref{conclusion}, where we also outline possible future directions.
\section{The Model}
\label{sec:model}
HNLs, also known as sterile neutrinos, represent a well-motivated extension of the SM, emerging naturally in several mechanisms for generating neutrino masses~\cite{Minkowski:1977sc,Yanagida:1979as,Gell-Mann:1979vob,Mohapatra:1979ia,Weinberg:1979sa,Schechter:1980gr,Foot:1988aq,Ma:1998dn,Asaka:2005pn,Bondarenko:2018ptm,Abdullahi:2022jlv, Han:2026ufa}. Beyond accounting for the smallness of active neutrino masses, HNLs may also explain the observed baryon asymmetry of the Universe through the mechanism of leptogenesis~\cite{Fukugita:1986hr,Fukugita:2002hu,Boyarsky:2009ix,Davidson:2008bu}.

HNLs are SM gauge singlets that couple to the SM lepton doublets via the so-called neutrino portal operator:
\begin{equation}
\mathcal{L} \supset Y_{\alpha} \, \bar{L}_\alpha (i\sigma^2 H^\ast) N + \text{h.c.},
\end{equation}
where \( Y_{\alpha} \) denotes the Yukawa coupling associated with the flavor of the lepton \( \alpha \), \( L_\alpha \) is the SM lepton doublet, and \( H \) is the Higgs doublet. After electroweak symmetry breaking, the Higgs acquires a vacuum expectation value \( v_\Phi \), generating a Dirac mass term for the neutrinos:
\begin{equation}
M_D = \frac{Y_{\alpha} v_\Phi}{\sqrt{2}}.
\end{equation}
If the HNLs also possess a Majorana mass term, they violate the lepton number by two units:
\begin{equation}
\mathcal{L} \supset -\frac{1}{2} M_N \, \bar{N}^c N + \text{h.c.}
\end{equation}
The complete mass terms, combining Dirac and Majorana contributions, lead to the following Type-I seesaw Lagrangian:
\begin{equation}
\mathcal{L}_{\text{Type-I}} = -\frac{1}{2}
\begin{pmatrix}
\overline{\nu_L} & \overline{N^c}
\end{pmatrix}
\begin{pmatrix}
0 & M_D \\
M_D^T & M_N
\end{pmatrix}
\begin{pmatrix}
\nu_L^c \\
N
\end{pmatrix} + \text{h.c.}
\end{equation}
Assuming a seesaw hierarchy \( M_N \gg M_D \), the effective mass matrix for light neutrinos is:
\begin{equation}
m_\nu^{\text{(I)}} = - M_D M_N^{-1} M_D^T.
\end{equation}

The flavor eigenstates of the light neutrinos can therefore be expressed as a superposition of three light-mass eigenstates $\nu_m$ and $n$ heavy-mass eigenstates $N_j$:
\begin{equation}
\nu_{L\ell} = \sum_{m=1}^3 P_{\ell m} \nu_m + \sum_{j=1}^n U_{\ell j} N_j, \quad \ell = e, \mu, \tau \, \, ,
\end{equation}
where $P_{\ell m}$ denotes the elements of the PMNS matrix describing active neutrino mixing, $U_{\ell j}$ characterizes the strength of active-sterile mixing, and $n$ is the number of HNLs. Although the heavy neutrino $N_j$ lacks direct gauge interactions, its coupling to the electroweak sector of the SM arises through this mixing. 
At leading order in the small mixing parameters $U_{\ell j}$, the charged-current interaction part of the Lagrangian, which is relevant for the present work, becomes:
\begin{equation}
\mathcal{L}_{\text{CC}} =
\frac{g}{\sqrt{2}} \sum_{\ell=e,\mu,\tau} W_\mu \bar{\ell}_L \gamma^\mu \left( \sum_{m=1}^{3} P_{\ell m} \nu_m +\sum_{j=1}^n U_{\ell j} N_j\right)  + \text{h.c.} \, \, ,
\end{equation}
where $g$ is the weak interaction coupling constant.

In the Type-I seesaw framework, the active-sterile mixing angle is typically given by \( U_{\ell j}^2 \sim m_\nu / M_N \). To obtain light neutrino masses of \( \mathcal{O}(0.1) \, \text{eV} \) with \( M_N \sim \mathcal{O}(1) \, \text{TeV} \), the Yukawa coupling must be tiny, \( Y_\alpha \sim \mathcal{O}(10^{-6}) \). If, instead, one assumes a coupling of \( Y_\alpha \sim \mathcal{O}(1) \), the HNL mass must be extremely large, well beyond the reach of current collider experiments.
This limitation has motivated the development of extended seesaw mechanisms that allow for lower HNL masses and larger mixing angles, making them potentially accessible in collider experiments such as the LHC. 
In this work, we adopt a phenomenological perspective that avoids reliance on specific high-scale realizations of the seesaw mechanism. Instead of embedding our analysis in a fully UV-complete model, we treat the active-sterile mixing angle $U_{\ell j}$ and the heavy neutrino mass $M_N$ as independent parameters. This approach is motivated by the desire to explore a broad and model-agnostic region of parameter space, capturing the key signatures of heavy neutrinos in a collider setting while remaining agnostic about the underlying new physics responsible for their origin.

In the high-mass region ($M_N \gtrsim 100\,\GeV$) which we are interested in, the direct LHC searches for the HNLs are limited by the collider energy (for more details see Ref.~\cite{Abdullahi:2022jlv, Granelli:2025lds} and references therein), and indirect constraints from EWPOs and charged lepton flavor violation (cLFV) remain the most stringent~\cite{delAguila:2008pw, Akhmedov:2013hec, deBlas:2013gla, Basso:2013jka, Antusch:2014woa, Antusch:2015mia, Chrzaszcz:2019inj, Bryman:2021teu, Blennow:2023mqx}. 
In ref.~\cite{Blennow:2023mqx}, an updated global fit analysis is performed by using new and updated experimental results on the key observables. 
Using the $Z$ boson mass, fine structure constant and Fermi constant extracted from muon decay as the SM input parameters in the electroweak sector, bounds on $\sum_j|U_{\ell j}|^2$ are derived by fitting to the $W$-boson mass $M_W$, the effective weak mixing angle $s_{\rm eff}^2$, LEP $Z$-pole observables ($\Gamma_Z$, $\sigma_{\rm had}^0$, $R_e$, $R_\mu$, $R_\tau$, $\Gamma_{\rm inv}$), tests of lepton-flavor universality in $\pi$, $K$, and $\tau$ decays, and weak-decay measurements constraining the unitarity of the Cabibbo-Kobayashi-Maskawa (CKM) matrix.
The global 95\% confidence-level (CL) limits on the mixing parameters are 
$\sum_j|U_{ej}|^2=[0.162,\,2.8]\times10^{-3}$, 
$\sum_j|U_{\mu j}|^2<2.8\times10^{-4}$, and 
$\sum_j|U_{\tau j}|^2<1.8\times 10^{-3}$ \cite{Blennow:2023mqx}.
The null observation of various cLFV processes provide constraints on the flavor-violating quantities $|\sum_j U_{\ell N} U^*_{\ell' N}|$ ($\ell\neq\ell'$). 
The current upper bounds from the flavor-violating observables are $|\sum_jU_{ej} U_{\mu j}^*|<2.4\times10^{-5}$, 
$|\sum_j U_{ej} U^*_{\tau j}|<1.6\times10^{-2}$, and 
$|\sum_j U_{\mu j} U^*_{\tau j}|<1.9\times 10^{-2}$, while 
the flavor-conserving observables, including the EWPOs, also put strong bounds on $|\sum_j U_{\ell N} U^*_{\ell' N}|$ through the Cauchy--Schwarz inequality, which leads to $|\sum_jU_{ej} U_{\mu j}^*|<6.8\times10^{-4}$, 
$|\sum_j U_{ej} U^*_{\tau j}|<1.8\times10^{-3}$, and 
$|\sum_j U_{\mu j} U^*_{\tau j}|<3.6\times 10^{-4}$~\cite{Blennow:2023mqx}.
\section{HNL searches at $\mu$TRISTAN}
\label{HNLsearch}
In this work, we investigate the potential for HNL searches in a promising proposed future collider: the $\mu^{+} \mu^{+}$ collider, also known as $\mu$TRISTAN. Due to the significantly lower synchrotron radiation emitted by muons compared to electrons, it is feasible to consider circular colliders operating at TeV-scale energies with circumferences of only a few kilometers~\cite{Kondo:2018rzx,Gallardo:1996aa,Ankenbrandt:1999cta}. Such a setup can also achieve sufficient luminosity for meaningful physics studies using existing technologies. While producing a narrow $\mu^{-}$ beam remains a technological challenge, there are well-established methods for generating low-emittance $\mu^{+}$ beams using ultra-cold muons~\cite{Kondo:2018rzx}, providing strong motivation to focus on a $\mu^{+} \mu^{+}$ collider. For this work, we consider the center-of-mass energy $\sqrt{s} = 10$~ TeV with integrated luminosity $\cal{L} =$ 1 ab$^{-1}$ and unpolarized $\mu^+$ beams.

\begin{figure}[t]
\centering
\begin{tikzpicture}
\begin{feynman}
    \vertex (mu1) {$\mu^+$};
    \vertex [below=3cm of mu1] (mu2) {$\mu^+$};
    \vertex [right=2.5cm of mu1] (v1);
    \vertex [right=2.5cm of mu2] (v2);
    \vertex [below=1.5cm of v1] (v3);
    \vertex [right=2cm of v1] (W+) {$W^+$};
    \vertex [right=2cm of v2] (vm) {$\bar{\nu}_\mu$};
    \vertex [right=2cm of v3] (tau) {$\tau^+$};
            
    \diagram* {
        (mu1) -- [anti fermion] (v1) -- [anti fermion, edge label=$N$] (v3) -- [anti fermion] (tau),
        (mu2) -- [anti fermion] (v2) -- [anti fermion] (vm),
        (v1) -- [photon] (W+),
        (v2) -- [photon, edge label=$W$] (v3),
        };
\end{feynman}
\end{tikzpicture}
\hspace{1cm}
\begin{tikzpicture}
\begin{feynman}
    \vertex (mu1) {$\mu^+$};
    \vertex [below=3cm of mu1] (mu2) {$\mu^+$};
    \vertex [right=2.5cm of mu1] (v1);
    \vertex [right=2.5cm of mu2] (v2);
    \vertex [below=1.5cm of v1] (v3) {};
    \vertex [right=2cm of v2] (v4) {$W^+$};
    \vertex [right=2cm of v1] (W+) {$W^+$};
            
    \diagram* {
        (mu1) -- [anti fermion] (v1) -- [majorana, edge label'=$N$] (v2) -- [fermion] (mu2),
        (v1) -- [photon] (W+),
        (v2) -- [photon] (v4),
        };
\end{feynman}
\end{tikzpicture}
\caption{Feynman diagrams for the signal processes: the LFV process  
$\mu^+\mu^+\to W^+ \tau^+ \bar{\nu}_\mu$ (left) and the LNV process
$\mu^+\mu^+\to W^+ W^+$ (right).
}
\label{fig:diagram}
\end{figure}
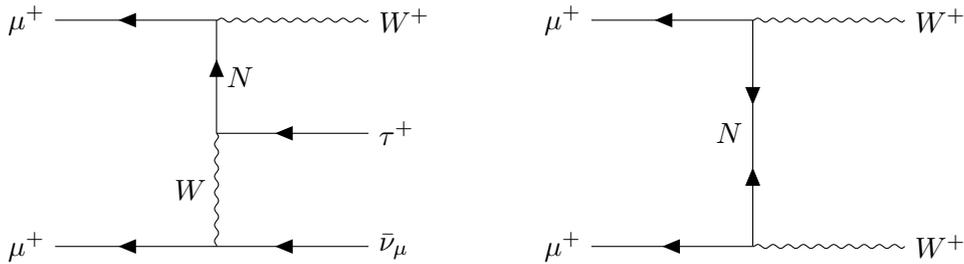
\subsection{Signal processes}
We focus on the following two HNL-mediated processes: (1) the LFV process $\mu^+\mu^+ \to W^+ \tau^+ \bar{\nu}_\mu$ and (2) the LNV process  $\mu^+\mu^+ \to W^+ W^+$. The corresponding Feynman diagrams are shown in Fig.~\ref{fig:diagram}.

The first process, $\mu^+\mu^+ \to W^+ \tau^+ \bar{\nu}_\mu$, violates the lepton flavor and provides a chance to access the $\nu_\tau$--$N$ mixing, $U_{\tau N}$.
The cross section is proportional to
\begin{align}
    \bigg|\sum_{I=1}^{\tilde n} U_{\mu I} U_{\tau I}^* \bigg|^2,
\end{align}
where $U_{\ell I}$ denotes the mixing between the active neutrino flavor
$\nu_\ell$ and the heavy mass eigenstate $N_I$
($I = 1, \dots, \tilde{n}$).
For simplicity, we assume that the \(\tilde{n}\) HNLs are nearly
degenerate in mass, while the remaining states are sufficiently heavy to be
kinematically irrelevant for the signal processes.
This LFV process does not rely on the Majorana nature of
neutrinos and can also arise in inverse-seesaw–type models.

The second process, $\mu^+\mu^+ \to W^+W^+$, is the muon--collider analog
of the inverse neutrinoless double beta decay ($0\nu\beta\beta$)
at $e^-e^-$ colliders, and it violates lepton number by two units.
The corresponding cross section is proportional to
\begin{align}
    \bigg|\sum_{I=1}^{\tilde n} U_{\mu I}^2 \bigg|^2.
\end{align}
As before, we assume that the \(\tilde n\) HNLs
labeled by $I$ are nearly degenerate in mass.
Since this process involves the Majorana mass term in the $t$-channel HNL
propagator, the cross section scales as $\sigma \propto M_N^2$ at high
$\sqrt{s}$, making it particularly sensitive to the high-mass region of
the HNL parameter space.
The parameters $\big|\sum_{I=1}^{\tilde n} U_{\mu I} U_{\tau I}^*\big|$
and $\big|\sum_{I=1}^{\tilde n} U_{\mu I}^2\big|$ are independent.
For instance, one can realize
$\big|\sum_{I=1}^{\tilde n} U_{\mu I} U_{\tau I}^*\big| = 0$
while keeping
$\big|\sum_{I=1}^{\tilde n} U_{\mu I}^2\big| \neq 0$
by taking $U_{\tau I} = 0$.
Conversely,
$\big|\sum_{I=1}^{\tilde n} U_{\mu I}^2\big| = 0$
can be achieved while maintaining
$\big|\sum_{I=1}^{\tilde n} U_{\mu I} U_{\tau I}^*\big| \neq 0$
if the Majorana mass term vanishes.

The EWPOs give the bound on the lepton-flavor-conserving quantities:
\begin{align}
\sum_{i=1}^n |U_{\ell i}|^2,
\label{UsqSum}
\end{align}
where $n$ is the number of the HNLs.
Note that the EWPOs bounds do not depend on the mass of the HNLs as long as they are much heavier than the mass of the $W$ boson.
The parameters $|\sum_{I=1}^{\tilde n} U_{\mu I}U_{\tau I}^*|$ and $|\sum_{I=1}^{\tilde n}U_{\mu I}^2|$ are bounded by combinations of Eq. (\ref{UsqSum}), using the Cauchy--Schwarz inequality
\begin{align}
   &\sum_{I=1}^{\tilde n} |U_{\mu I}|^2 \sum_{J=1}^{\tilde n} |U_{\tau J}|^2 - \left|\sum_{I=1}^{\tilde n} U_{\mu I} U_{\tau I}^*\right|^2 =\sum_{I>J}|U_{\mu I}U_{\tau J} - U_{\mu J}U_{\tau I}|^2\geq 0,\\
   &\left(\sum_{I=1}^{\tilde n} |U_{\mu I}|^2\right)^2 - \left|\sum_{I=1}^{\tilde n} U_{\mu I}^2\right|^2 =\sum_{I>J}|U_{\mu I}U_{\mu J}^* - U_{\mu J}U_{\mu I}^*|^2\geq 0,
\end{align}
and the trivial inequality,
\begin{align}
    \sum_{i=1}^n|U_{\ell i}|^2 \geq \sum_{I=1}^{\tilde n}|U_{\ell I}|^2.
\end{align}
Therefore, the bounds from EWPOs on the quantities Eq.~(\ref{UsqSum}) also constrain the parameters $\left|\sum_{I=1}^{\tilde n} U_{\mu I}U_{\tau I}^*\right|$ and $|\sum_{I=1}^{\tilde n}U_{\mu I}^2|$.
Although an upper bound on
\( \big|\sum_{i=1}^{n} U_{\mu i} U_{\tau i}^*\big| \)
is discussed in Sec.~\ref{sec:model}, it does not directly apply to
\( \big|\sum_{I=1}^{\tilde n} U_{\mu I} U_{\tau I}^*\big| \),
as cancellations involving contributions from heavier neutrinos may occur.

\begin{figure}[t]
    \centering
    \includegraphics[width=0.6\textwidth]{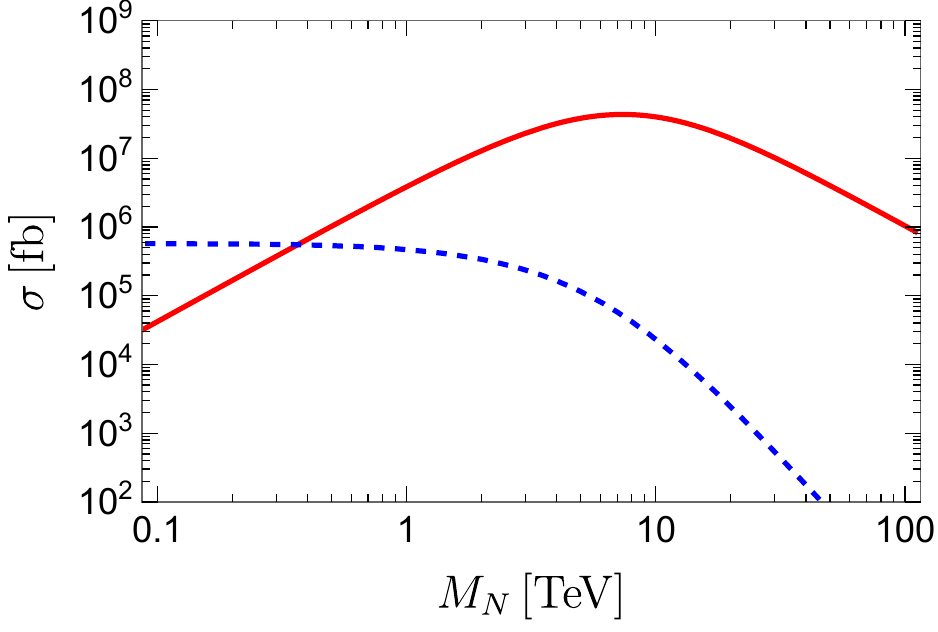}
    \caption{The cross sections for $\mu^+\mu^+\to W^+\tau^+\bar{\nu}_\mu$ (blue dashed) and $\mu^+\mu^+\to W^+W^+$ (red) at $\sqrt{s}=10\,{\rm TeV}$, with $|\sum_{I=1}^{\tilde n} U_{\mu I}U_{\tau I}^*|=|\sum_{I=1}^{\tilde n} U_{\mu I}^2|=1$.}
    \label{fig:xsec}
\end{figure}
The inverse $0\nu\beta\beta$ process at same-sign lepton colliders has been studied in the literature as a high-energy probe of Majorana nature of neutrinos \cite{London:1987nz, Belanger:1995nh, Heusch:1993qu, Heusch:1995yw, Gluza:1995ix, Asaka:2015oia, Wang:2016eln, Han:2026ufa}.  
The full cross section in the presence of a heavy Majorana neutrino is analytically computed in \cite{London:1987nz, Belanger:1995nh}. 
For our purpose, it is useful to take the $m_W\ll\sqrt{s}$ limit, finding an approximate cross section~\cite{Asaka:2015oia}, 
\begin{equation}
\sigma_{W^+W^+} \simeq 
\frac{G_F^2 s |\sum_{I=1}^{\tilde n}U_{\mu I}^2|^2}{8\pi} H(M_N^2/s) \ ,
\label{eq:xsec_WW}
\end{equation}
where $U_{\mu I}$ denotes the $\nu_\mu$--$N_I$ mixing and 
\begin{equation}
H(r)=r(2+3r)\left[\frac{1}{1+r}-\frac{2r}{1+2r}\log\left(1+\frac{1}{r}\right)\right] \ ,
\end{equation}
which takes the maximum value, $H(r)_{\rm max}\simeq0.21$, at $r=0.54$ (i.e. $M_N\simeq0.74\sqrt{s}$).
In the high-energy limit ($\sqrt{s}\gg M_N$) and the heavy HNL limit ($\sqrt{s}\ll M_N$), it further approximates to 
\begin{equation}
\renewcommand{\arraystretch}{2}
\sigma_{W^+W^+} \simeq 
\left\{\begin{array}{ll}
    \displaystyle \frac{G_F^2 M_N^2}{4\pi}\displaystyle \left|\sum_{I=1}^{\tilde n}U_{\mu I}^2 \right|^2 & ~~\mbox{for $\sqrt{s}\gg M_N$}\\
    \displaystyle \frac{\displaystyle G_F^2 s^2}{16\pi M_N^2} \left|\sum_{I=1}^{\tilde n}U_{\mu I}^2 \right|^2  & ~~\mbox{for $\sqrt{s}\ll M_N$}
\end{array}\right..
\label{eq:xsec_WW_approx}
\renewcommand{\arraystretch}{1}
\end{equation}
Note here that the cross section vanishes at $M_N = 0$ for fixed $\sqrt s$, and scales as $1/M_N^2$
for large $M_N$. This can be understood by the fact that the process needs LNV so that amplitude is proportional to $M_N$ or $1/M_N$ for small or large $M_N$, respectively.
In contrast, the amplitude of the LFV process does not require an $M_N$ insertion. 
Therefore, it is expected to scale as $\mathcal{O}(M_N^0)$ for $M_N \ll \sqrt{s}$, 
while for $M_N \gg \sqrt{s}$ the amplitude scales as $1/M_N$, 
leading to a cross section proportional to $1/M_N^2$.

One can see such qualitative differences in Fig.~\ref{fig:xsec} where the $M_N$ dependence of the signal cross sections at $\mu^+\mu^+$ colliders is shown. The parameters are chosen as $\sqrt{s}=10\,{\rm TeV}$ and $|\sum_{I=1}^{\tilde n} U_{\mu I}U_{\tau I}^*|=|\sum_{I=1}^{\tilde n} U_{\mu I}^2|=1$. 
The red line corresponds to the cross section of $\mu^+\mu^+ \to W^+ W^+$, which we calculate using 
Eq.~(\ref{eq:xsec_WW}). 
As suggested by the analytical approximation Eq.~(\ref{eq:xsec_WW_approx}) in two limiting cases, the cross section is maximized at $M_N \simeq \sqrt{s}$. 
The blue dashed line shows the cross section of $\mu^+\mu^+ \to W^+ \tau^+ \bar{\nu}_\mu$, which we calculate numerically with {\tt MadGraph5\_aMC@NLO} \cite{Alwall:2014hca}.
At the boundary of the electroweak precision bounds \cite{Blennow:2023mqx}, i.e. $
|\sum_{I=1}^{\tilde n} U_{\mu I}U_{\tau I}^*|\simeq7.1\times10^{-4}$ and $|\sum_{I=1}^{\tilde n}U_{\mu I}^2|\simeq2.8\times10^{-4}$, 
we have 
$\sigma_{W^+\tau^+\bar{\nu}_\mu}\simeq0.06\,{\rm fb}$ and 
$\sigma_{W^+W^+}\simeq0.3\,{\rm fb}\,(M_N/{\rm TeV})^2$ for $\sqrt{s}\gg M_N$.
Thus, 300 or more $W^+W^+$ events are expected at $\sqrt{s}=10\,{\rm TeV}$ with $1\,{\rm ab}^{-1}$ luminosity.
\subsection{Simulation and analysis}
In this section, we estimate the expected sensitivities to the HNL mass and to the mixing combinations $|\sum_{I=1}^{\tilde n} U_{\mu I}U_{\tau I}^*|$ and $|\sum_{I=1}^{\tilde n} U_{\mu I}^2|$ at a same-sign muon collider with $\sqrt{s}=10~\mathrm{TeV}$ and an integrated luminosity of $1~\mathrm{ab}^{-1}$. As stated above, we consider the two signal events, the LNF process $\mu^+\mu^+\to W^+\tau^+\bar\nu_\mu$ and the LNV process $\mu^+\mu^+\to W^+W^+$. 
The final states are analyzed separately according to the decay modes of the $W^+$ boson and $\tau$ lepton. 
After optimizing the selection cuts on the relevant kinematic variables to maximize signal–background discrimination, we find that the singlet-$\tau$ plus non-$\tau$ jet
topology provides the highest sensitivity for the LFV process, whereas the fully hadronic two-jet topology offers the greatest sensitivity for the LNV process. Contributions from the remaining channels are therefore neglected in the following analysis.
Signal and background yields for these target final states are obtained from numerical simulations.
We apply event selections designed to enhance the signal significance and derive the expected limits under the background-only hypothesis, i.e., assuming that the observed number of events equals the expected background yield.

Both signal and background events are generated using {\tt MadGraph5\_aMC@NLO} \cite{Alwall:2014hca}. 
To generate the signal events MadGraph5 is interfaced with the {\tt SM\_HeavyN\_NLO} model file as obtained from the package FeynRules Model Database \cite{Alloul:2013bka}.  
The decay kinematics of unstable particles, as well as subsequent parton showering and hadronization, are simulated with {\tt Pythia8} \cite{Sjostrand:2006za,Sjostrand:2007gs}.
Finally, detector responses are taken into account using {\tt Delphes} \cite{deFavereau:2013fsa} employing the MuonCollider card with the {\tt VLCjetR05N2} Jet Branch.

\begin{table}[t]
\centering
\vspace{10pt}
\begin{tabular}{c|c|c|c|c|c}
\hline
$M_N$ [TeV]                             &0.1  &0.5  &1   &2    &4    \\\hline
$|\sum_{I=1}^{\tilde n} U_{\mu I}U_{\tau I}^*| \times 10^{4}$           &6.1  &6.3  &6.3 &6.3  &8.9  \\\hline
$p_T^{\mathrm{\tau jet}}>x$ [TeV]    &0.8  &0.7  &0.7 &0.6  &0.5  \\\hline
$p_T^{\mathrm{non\tau jet}}>x$ [TeV] &3.0  &3.0  &2.9 &3.2  &3.2  \\\hline
$M^{\mathrm{tot}}>x$ [TeV]           &3.2  &3.0  &2.7 &0.0  &3.0  \\\hline
$M^{\mathrm{non\tau jet}}<x$ [GeV]   &88.6 &88.4 &88.5&88.8 &88.6 \\\hline
\end{tabular}
\caption{Optimized selection cuts for the LFV process at five benchmark values of $M_N$. Analogous selection cuts for the LNV process are discussed in the text.}
\label{tab1}
\end{table}

\begin{table}[t]
\centering
\vspace{10pt}
\begin{tabular}{c|c|c|c}
\hline
     &Signal &Background & $S/\sqrt{S+B}$\\\hline\hline
Total                                               &$181$  & $6.59\times10^6$ & $7.06\times10^{-2}$ \\\hline
Exact 1 $\tau$-jet                                  &$75.8$ & $3.57\times10^5$ & $0.127$ \\\hline
$\tau$-jet has charge 1                             &$74.6$ & $2.04\times10^5$ & $0.165$ \\\hline
At least 1 non-$\tau$-jet                           &$74.6$ & $2.04\times10^5$ & $0.165$ \\\hline
no electrons and muons                              &$63.1$ & $1.59\times10^5$ & $0.158$ \\\hline
$p_T^{\tau\text{-}\mathrm{jet}}>0.7$ TeV            &$47.2$ & $2.76\times10^3$ & $0.89$              \\\hline
$p_T^{\text{non-}\tau\text{-}\mathrm{jet}}>2.9$ TeV &$12.7$ & $25.5$           & $2.06$              \\\hline
$M^{\mathrm{tot}}>2.7$ TeV                          &$12.7$ & $25.4$           & $2.06$              \\\hline
$M^{\text{non-}\tau\text{-jet}}>88.5$ GeV           &$11.3$ & $13.8$           & $2.27$              \\\hline
\end{tabular}
\caption{Number of events after each cuts for the LFV process at $|\sum_{I=1}^{\tilde n} U_{\mu I}U_{\tau I}^*| = 6.3\times 10^{-4}$ and $M_N=1$ TeV. }
\label{tab2}
\end{table}

\begin{table}[t]
\centering
\vspace{10pt}
\begin{tabular}{c|c|c|c}
\hline
     &Signal &Background & $S/\sqrt{S+B}$\\\hline\hline
Total                            &$16.3$ & $6.93\times10^6$ & $6.19\times10^{-2}$ \\\hline
At least 2 jets                  &$15.2$ & $4.11\times10^6$ & $7.50\times10^{-2}$ \\\hline
No electrons and muons           &$9.64$ & $2.99\times10^6$ & $5.57\times10^{-2}$ \\\hline
$p_T^{\mathrm{jet}1}>4.2$ TeV    &$4.47$ & $7.12$           & $1.31$              \\\hline
$p_T^{\mathrm{jet}2}>1.9$ TeV    &$3.92$ & $1.63$           & $1.66$              \\\hline
$M^{\mathrm{tot}}>6.45$ TeV      &$3.87$ & $1.47$           & $1.67$              \\\hline
\end{tabular}
\caption{Number of events after each cuts for the LNV process at $|\sum_{I=1}^{\tilde n}U_{\mu I}^2|=6.5\times10^{-5}$ and $M_N=1$ TeV.}
\label{tab3}
\end{table}
We simulate the following relevant SM background processes for our analysis:
(i) $\mu^+\mu^+\to \mu^+\mu^+ Z$,
(ii) $\mu^+\mu^+ \to \mu^+\mu^+ ZZ$,
(iii) $\mu^+\mu^+ \to \mu^+\mu^+ W^+ W^-$,
(iv) $\mu^+\mu^+ \to \mu^+\bar\nu_\mu Z W^+$, and
(v) $\mu^+\mu^+ \to \bar\nu_\mu\bar\nu_\mu W^+W^+$.
In addition, we also consider the processes $\mu^+\mu^+ \to \mu^+\mu^+\tau^+\tau^-$ and $\mu^+\mu^+ \to \mu^+\mu^+ Z \tau^+\tau^-$, which overlap with the backgrounds (i) and (ii) when the $Z$ boson decays into $\tau^+\tau^-$.
In our analysis, contributions from these backgrounds to the signal region are largely suppressed by imposing a cut on the transverse momentum ($p_T$) of jets, and therefore we neglect these two processes in the following analysis\footnote{
After introducing the cuts described in Tab.~2 in the draft, the expected event number is less than $1.19\times10^{-3}$ for $\mu^+\mu^+\to\mu^+\mu^+\tau^+\tau^-$ and less than $4.5\times10^{-2}$ for $\mu^+\mu^+\to\mu^+\mu^+Z\tau^+\tau^-$.
}.

The background samples for (iv) and (v) are generated with $10^6$. Fewer events are generated for (i) and (ii), since they do not contribute to the signal region. For background process (v), we generate $3\times10^6$ events to ensure sufficient statistical precision.~\footnote{
We have checked that repeating the background simulation with three statistically independent samples, each containing $10^6$
  events, produces consistent projected sensitivities within the expected Monte Carlo statistical fluctuations. This indicates that the statistical uncertainty associated with the finite size of the simulated background sample is negligible for the results presented here.
} Signal samples are generated with $10^5$ events.
The cross sections for processes with $\mu^+$ in the final state suffer from divergences due to collinear photon emission when evaluated using \texttt{MadGraph}.
For these processes, when applicable, we compute the total cross sections using the method described in Ref.~\cite{Hamada:2024ojj}, which splits the phase space into low- and high-$p_T$ regions. 
In Appendix \ref{app:ptcut}, we show the calculation of the background (iv) by using the method.
For backgrounds (iii) and (iv), we find suitable minimum $p_T$ cuts of $0.02$~GeV and $0.6$~GeV, respectively. For backgrounds (i) and (ii), this procedure is not applicable; instead, we adopt several minimum $p_T$ cuts in the range
$0.01~\mathrm{GeV} < p_T < 0.1~\mathrm{GeV}$.
Finally, we confirm that the jet $p_T$ cut applied in the analysis effectively removes these backgrounds, and therefore no further treatment of these processes is required.

The LFV and LNV signal yields are parameterized by the mixing combinations,
\[
\kappa_1 \equiv \Bigl|\sum_{I=1}^{\tilde n} U_{\mu I}U_{\tau I}^*\Bigr|, 
\qquad
\kappa_2 \equiv \Bigl|\sum_{I=1}^{\tilde n} U_{\mu I}^2\Bigr|.
\]
For each channel, a reference point is defined, $\kappa_1=c_1$ (LFV) and $\kappa_2=c_2$ (LNV), and the corresponding expected signal yields $S_1$ and $S_2$ are obtained from simulation under a fixed selection. Since the rates scale with the square of the mixing, the expected signal at generic $\kappa_i$ is
\[
S_1(\kappa_1) = S_1 \left(\frac{\kappa_1}{c_1}\right)^2, 
\qquad
S_2(\kappa_2) = S_2 \left(\frac{\kappa_2}{c_2}\right)^2.
\]
\begin{figure}[t]
    \centering
    \includegraphics[width=0.49\textwidth]{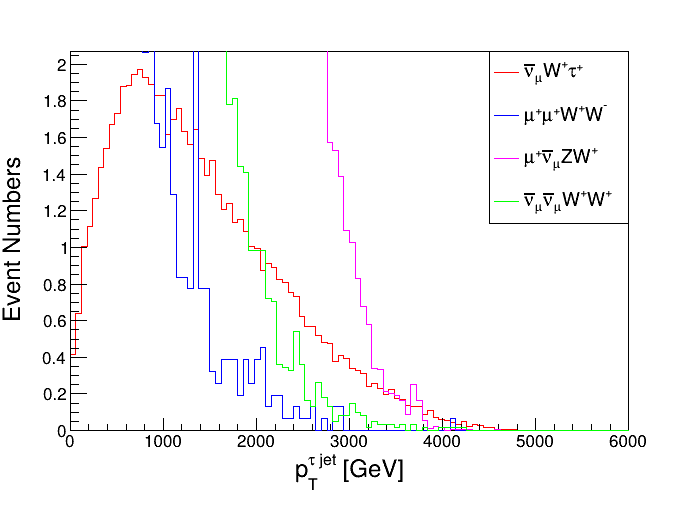}
    \includegraphics[width=0.49\textwidth]{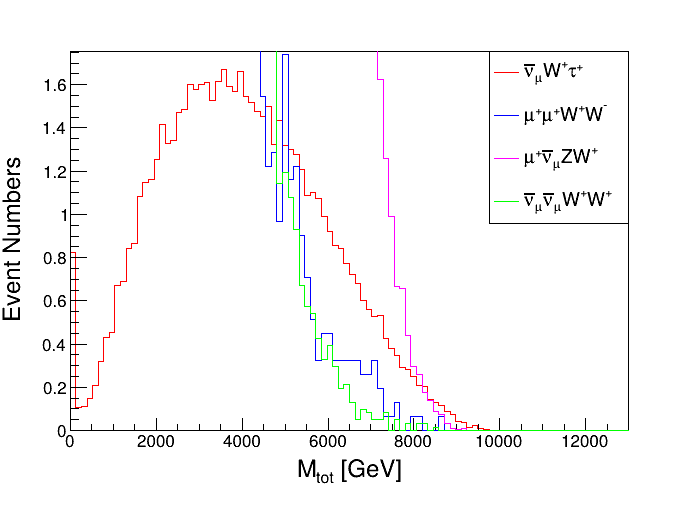}
    \includegraphics[width=0.49\textwidth]{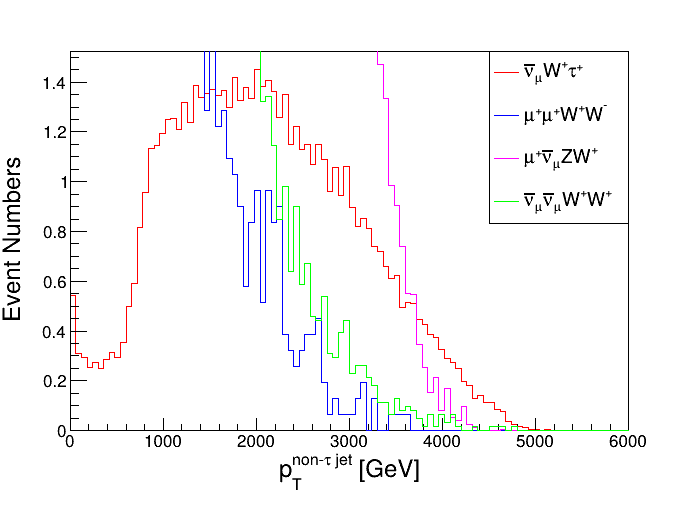}
    \includegraphics[width=0.49\textwidth]{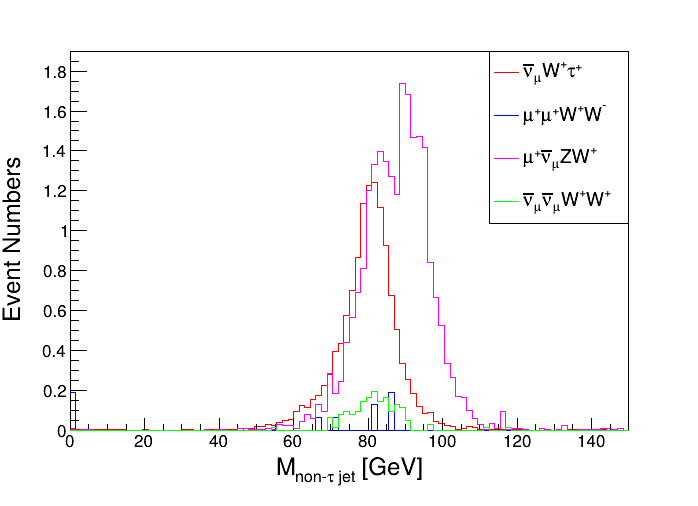}
\caption{Distributions of various kinematic observables for the individual SM background processes and the LFV signal channel corresponding to $|\sum_{I=1}^{\tilde n} U_{\mu I}U_{\tau I}^*| = 6.3\times10^{-4}$. The panels correspond to the observables (top-left) $P_T^{\tau \, \text{jet}}$, (top-right) $M_{\text{tot}}$, (bottom-left) $P_T^{\text{non-}\tau \, \text{jet}}$, and (bottom-right) $M_{\text{non-}\tau \, \text{jet}}$, all in GeV. The distribution of $M_{\text{non-}\tau \, \text{jet}}$ is shown after applying the cuts for $p^{\tau-\text{jet}}$, $M_{\text{tot}}$ and $p_T^{\text{non-}\tau\text{-jet}}$.}
\label{hist1}
\end{figure}
Separate 95\%\,CL upper limits on $\kappa_1$ (LFV) and $\kappa_2$ (LNV) are set, assuming the other combination vanishes in each case.
Let $B_1$ and $B_2$ denote the expected background yields after selection, and let $N_1$ and $N_2$ be the observed event counts.
In the expected-limit treatment, $N_i$ is taken to be the integer nearest to $B_i$. The 95\%\,CL upper limits are obtained by solving
\begin{align}
S_1 \left(\frac{\kappa_1}{c_1}\right)^2 + B_1 &= C_u(N_1), \nonumber\\
S_2 \left(\frac{\kappa_2}{c_2}\right)^2 + B_2 &= C_u(N_2),
\end{align}
where $C_u(N)$ denotes the upper endpoint of the exact two-sided
$95\%$ confidence interval for the mean of a Poisson distribution.
Explicitly, $C_u(N)$ is defined by
\begin{equation}
\sum_{k=0}^{N}
\frac{e^{-C_u(N)}[C_u(N)]^k}{k!}
=0.025.
\end{equation}
For example, $C_u(0)=3.688879$ and $C_u(1)=5.5716$.
Here, $C_u(N_i)$ is the confidence bound on the total Poisson
mean $S_i+B_i$. For a fixed background expectation $B_i$, the
corresponding upper limit on the signal yield is therefore
$S_i^{95}=C_u(N_i)-B_i$.

\begin{figure}[t]
    \centering
    \includegraphics[width=0.512\textwidth]{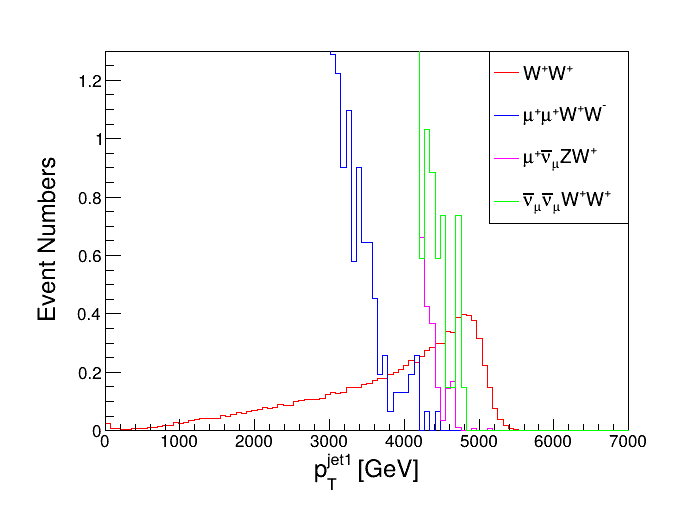}
    \includegraphics[width=0.491\textwidth]{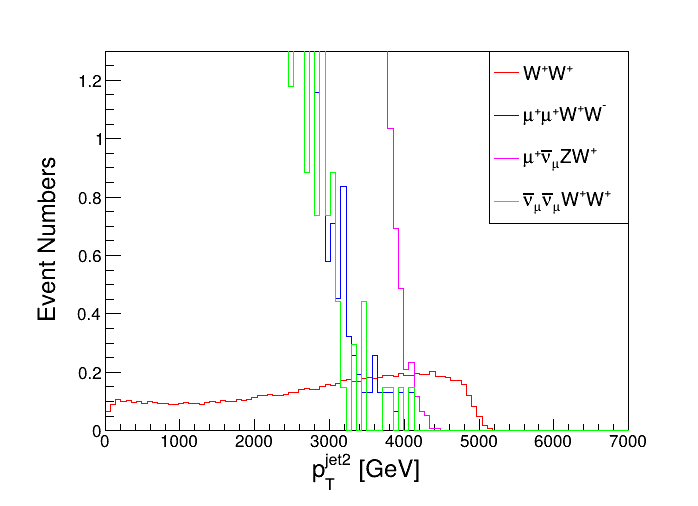}
    \includegraphics[width=0.491\textwidth]{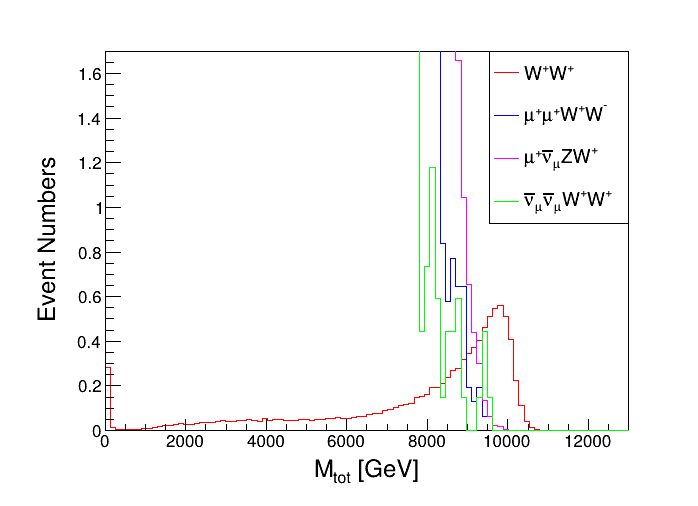}
\caption{Distributions of various kinematic observables (top) $P_T^{\text{jet1}}$, (bottom-left) $P_T^{\text{jet2}}$ and (bottom-right) $M_{\text{tot}}$, all in GeV, for the individual SM background processes and the LNV signal channel corresponding to $|\sum_{I=1}^{\tilde n}U_{\mu I}^2|  = 6.5\times10^{-5}$, $M_N=1$~TeV.}
\label{hist2}
\end{figure}

Selections are chosen to approximately maximize the expected significance $S/\sqrt{S+B}$. For the LFV channel, the cuts listed in Tab.~\ref{tab1} are used. For the LNV channel, the cuts are
\[
p_T^{\mathrm{jet}1}>4.2~\text{TeV},\quad
p_T^{\mathrm{jet}2}>1.9~\text{TeV},\quad
M_{\mathrm{tot}}>6.45~\text{TeV}\ \text{or}\ 6.75~\text{TeV},
\]
where $M_{\mathrm{tot}}$ is the invariant mass of the two jets, and we choose one of the two values of the $M_{\mathrm{tot}}$ threshold in the equation for each $M_N$.
An example cut flow table for the LFV (LNV) process considering various optimized cuts for the choice of mixing parameters $|\sum_{I=1}^{\tilde n} U_{\mu I}U_{\tau I}^*| = 6.3\times 10^{-4}$ ($|\sum_{I=1}^{\tilde n}U_{\mu I}^2|=6.5\times10^{-5}$) and $M_N=1$~TeV is shown in Tab.~\ref{tab2}~(\ref{tab3}).
Our choices of optimized cuts are guided by the distributions in Figs.~\ref{hist1} and \ref{hist2} for the LFV and LNV processes, respectively.

\begin{figure}[t]
\centering
\includegraphics[width=0.49\hsize]{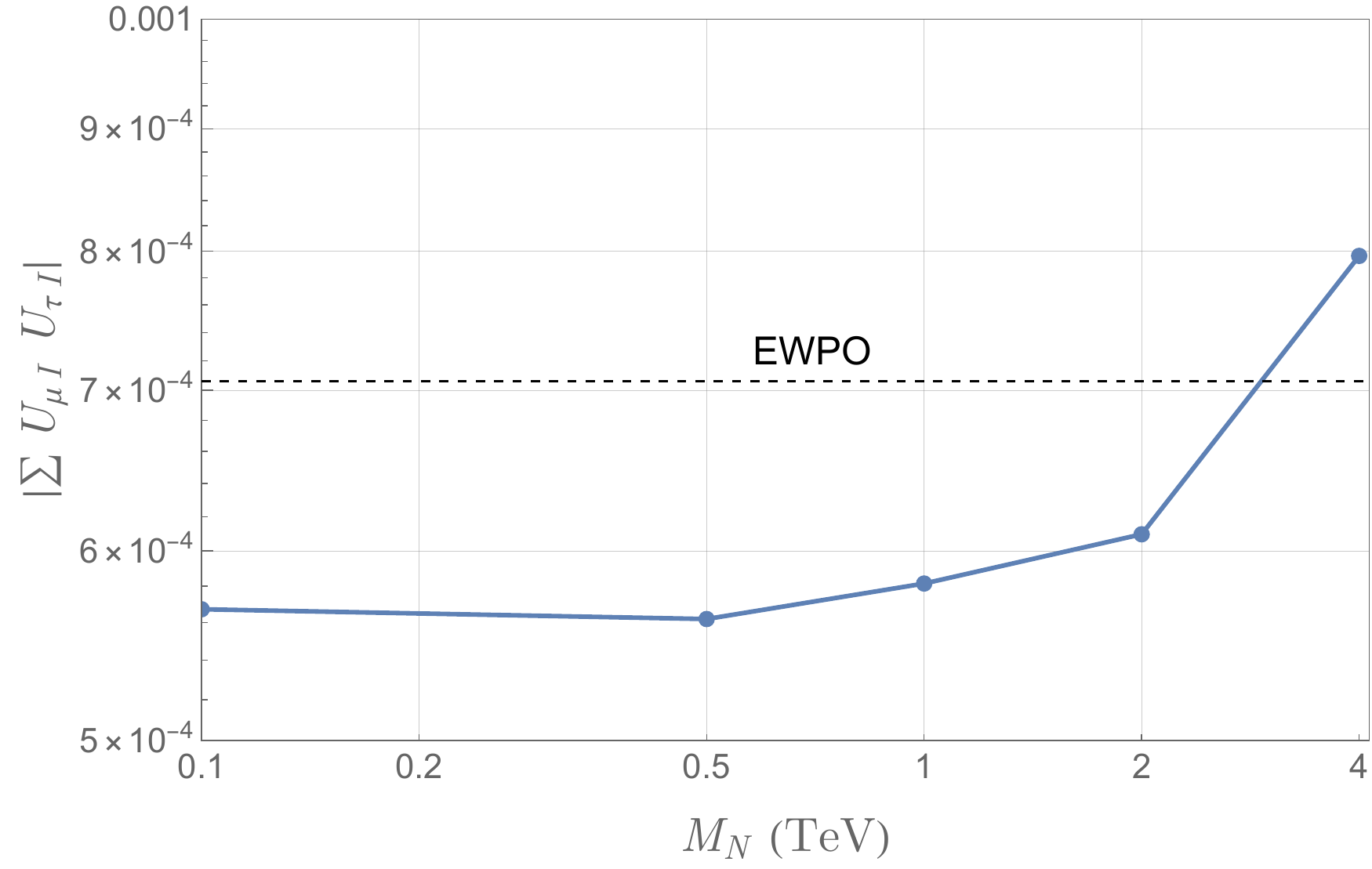}~~~~~~
\includegraphics[width=0.49\hsize]{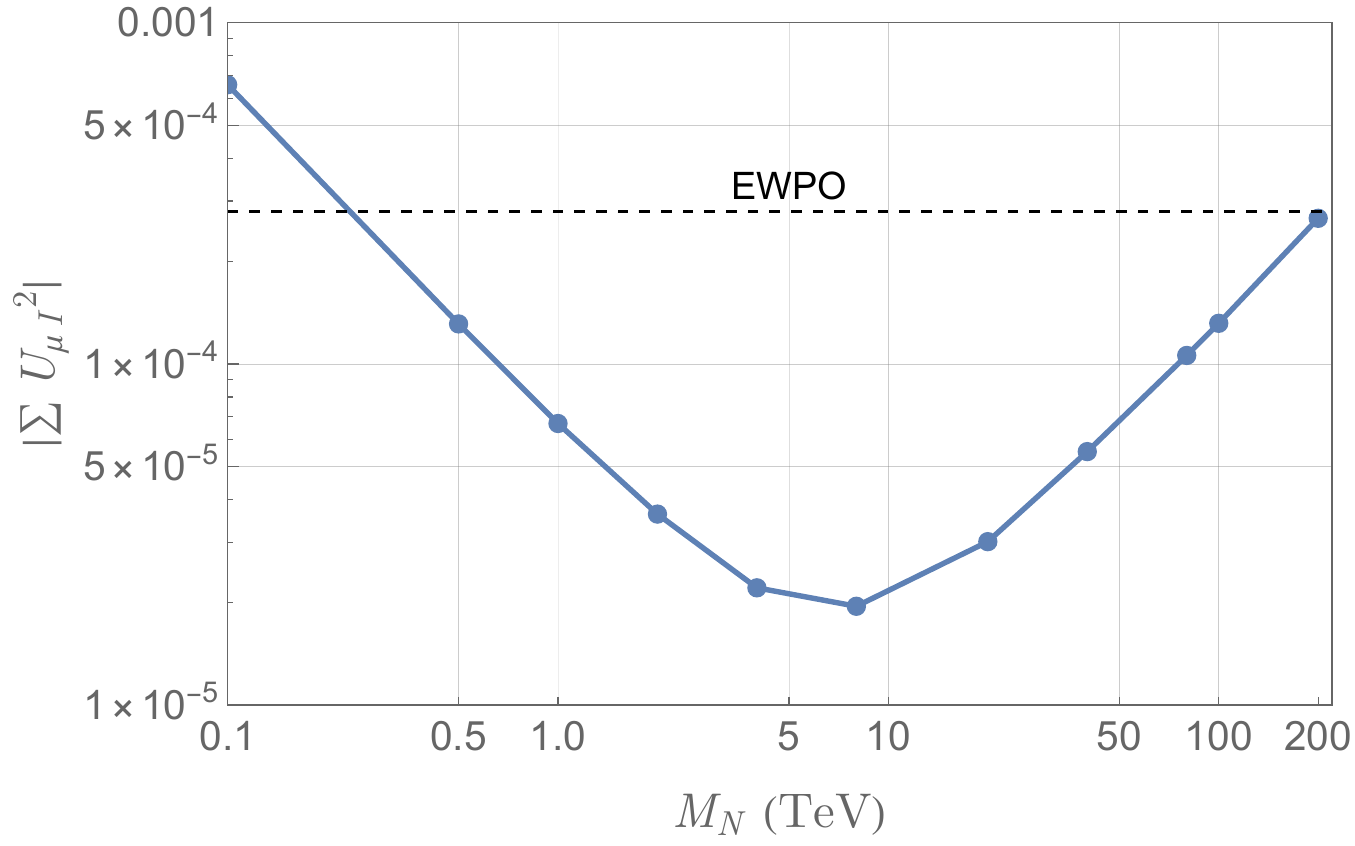}
\caption{The projected sensitivity (blue lines) in the $M_N-|\sum_{I=1}^{\tilde n} U_{\mu I}U_{\tau I}|$ (left) and $M_N-|\sum_{I=1}^{\tilde n} U_{\mu I}^2|$ (right)  planes. The current exclusion limit from the measurements of the EWPOs is indicated by the black dashed line.  
}
\label{bound}
\end{figure}

Scanning over $M_N$, we obtain the projected sensitivities of our analyses. 
Figure~\ref{bound} summarizes the reach: the left panel shows the $(M_N,\ |\sum_{I=1}^{\tilde n} U_{\mu I}U_{\tau I}^*|)$ plane (LFV), and the right panel shows the $(M_N,\ |\sum_{I=1}^{\tilde n} U_{\mu I}^{2}|)$ plane (LNV). 
In both cases, the expected sensitivity surpasses current bounds from the measurements of EWPOs over sizable mass ranges. 
For example, in the LFV case (left), our analysis is stronger than EWPO for $M_N \lesssim 3$~TeV. 
In the LNV case (right), the reach improves the constraint on $|\sum_{I=1}^{\tilde n} U_{\mu I}^{2}|$ by roughly an order of magnitude around $M_N \simeq 5$-$10$~TeV.
Before concluding, we emphasize that this work explores searches for HNL in two complementary channels using purely cut-based selections. To fully exploit the discovery reach of $\mu$TRISTAN, a natural next step is to develop machine-learning–based multivariate analyses (e.g., BDTs or neural networks) built on an expanded set of kinematic observables; such methods are expected to deliver substantially improved sensitivity relative to the results reported here.

\section{Conclusions}
\label{conclusion}
In this work we have explored the potential of a same-sign muon collider ($\mu$TRISTAN) to probe HNLs and their possible Majorana nature. Our study focused on two complementary processes, supported by detailed Monte Carlo simulations and background analyses:

\begin{itemize}

  \item LFV channel: $\boldsymbol{\mu^+\mu^+ \to W^+\tau^+\bar\nu_\mu}$.
  This process probes the flavor structure through the mixing combination $|\sum_I U_{\mu I}U^{*}_{\tau I}|$ and does not require lepton–number violation.
  We found the most sensitive final state to be one $\tau$–jet plus one non–$\tau$ jet.
  Key selections include a hard $\tau$–jet and non–$\tau$ jet transverse momentum and moderate global mass requirements, e.g.
  $p_T(\tau\text{-jet}) \gtrsim 0.5\text{–}0.8~\mathrm{TeV}$ (mass–dependent),
  $p_T(\text{non-}\tau\text{ jet}) \!\sim\! 3.0\text{–}3.2~\mathrm{TeV}$,
  $M_{\mathrm{tot}} \gtrsim 2.7\text{–}3.2~\mathrm{TeV}$,
  and an on–shell $W$ window for the non–$\tau$ jet ($m_{\text{non-}\tau\text{ jet}}\!\approx\! m_W$).
  With these cuts and at $\sqrt{s} = 10~\text{TeV}$ and $1~\text{ab}^{-1}$, the projected sensitivity curve
  in the $(M_N,\ |\sum_I U_{\mu I}U^{*}_{\tau I}|)$ plane improves upon the current bounds coming from the EWPOs across a broad $M_N$ range.

  \item \textbf{LNV channel: $\boldsymbol{\mu^+\mu^+ \to W^+W^+}$.}
  This process directly probes LNV via a Majorana neutrino propagator and depends on $|\sum_I U_{\mu I}^2|$.
  We identified the fully hadronic two–jet topology as optimal, with stringent requirements on the reconstructed $W$ candidates:
  $p_T(j_1) > 4.2~\mathrm{TeV}$, $p_T(j_2) > 1.9~\mathrm{TeV}$, and $M_{\mathrm{tot}} > 6.45\text{–}6.75~\mathrm{TeV}$ (mass–dependent).
  In the high–energy regime, the cross section follows the known scaling $\sigma(\mu^+\mu^+\!\to\!W^+W^+)\propto G_F^2\,s\,\big|\sum_I U_{\mu I}^2\big|^2\,H(M_N^2/s)$, peaking near $M_N\simeq0.7\sqrt{s}$, and yields $\mathcal{O}(10^2)$ signal events at benchmark mixings close to the current precision limits.
  The resulting reach in $(M_N,\ |\sum_I U_{\mu I}^2|)$ surpasses the existing indirect constraints over wide mass ranges.

\end{itemize}
The two channels probe distinct aspects of the HNL sector:
  the LFV process provides direct sensitivity to the flavor structure of active–sterile mixing (independent of whether neutrinos are Majorana),
  whereas the LNV process directly tests LNV and the Majorana nature.
  Taken together, they offer complementary discovery handles and a nontrivial consistency check on neutrino–mass generation scenarios.
It is also interesting to note that two processes
we considered involve initial muons with right-handed helicities, $\mu^+_R \mu^+_R$, as they need to couple to $W$ bosons. If the polarization of the $\mu^+$ beam is available,
the polarization dependence of the number of signals
would be a great handle to confirm the excess over the
background events. In addition, such analyses can discriminate
the underlying models. For example, the model discussed
in Ref.~\cite{Harigaya:2025zru} gives a similar signal when the initial state polarization is $\mu^+_L \mu^+_R$.  

Overall, our results demonstrate that $\mu$TRISTAN would significantly extend the experimental sensitivity to HNLs, opening new discovery opportunities beyond the HL-LHC and future $e^+e^-$ machines, and providing a decisive test of whether neutrinos are Majorana particles.

\section*{Acknowledgements}
This work is supported in part by the JSPS KAKENHI Grant Numbers 22K21350 (R.K., R.M., and S.O.) and 25K17401 (S.O.) and
the U.S.-Japan Science and Technology Cooperation Program in High Energy Physics~2025-20-2~(R.K., R.M., S.O., and I.L.).
S.O. is also supported by an appointment to the JRG Program at the APCTP through the Science and Technology Promotion Fund and Lottery Fund of the Korean Government and by the Korean Local Governments -- Gyeongsangbuk-do Province and Pohang City.
I.L. and S.R. are supported by the U.S.~Department of Energy under contracts No.\ DEAC02-06CH11357 at Argonne National Laboratory. I.L. is also supported by the U.S.~Department of Energy, Office of High Energy Physics, under contract No. DE-SC0010143. S.R. would like to thank the University of Chicago and Fermilab, Aspen Center for Physics and Perimeter Institute for Theoretical Physics, where a significant part of this work has been done.

\appendix
\section{Evaluation of the background with the collinear divergence}
\label{app:ptcut}
\begin{figure}[t]
\centering
\includegraphics[width=0.5\hsize]{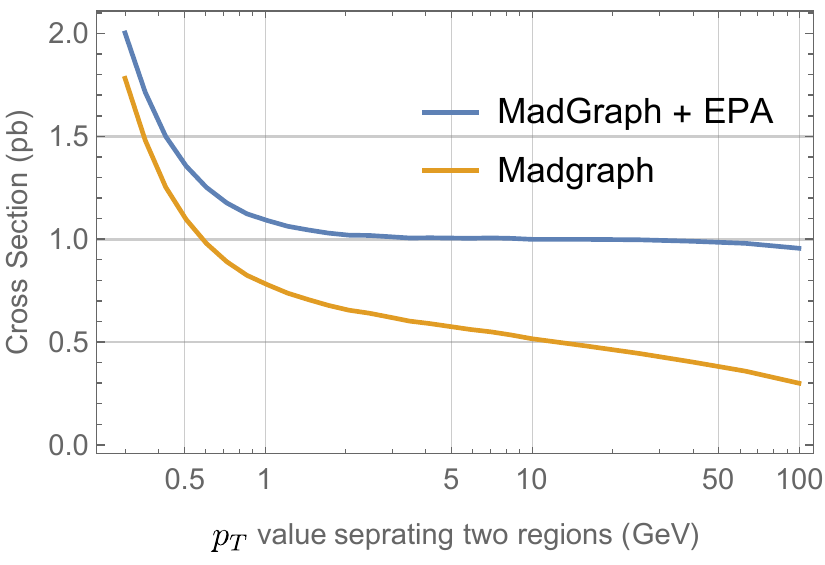}
\caption{The dependence on the $p_T$ value separating the two regions of the sum of the \texttt{MadGraph} and the EPA contributions, as well as the \texttt{MadGraph} contribution for the background process $\mu^+ \mu^+ \to \mu^+ \bar\nu_\mu W^+Z$.}
\label{fig:ptcut}
\end{figure}

In this appendix, we demonstrate how to obtain the correct cross section while avoiding the collinear divergence by using the method described in Ref.~~\cite{Hamada:2024ojj}.
In this approach, we split the phases space into low- and high-$p_T$ regions. 
In the low-$p_T$ region, the cross section is evaluated using the effective photon approximation~\cite{vonWeizsacker:1934nji, Williams:1934ad}, while in the high-$p_T$ region it is computed numerically with \texttt{MadGraph}. The two contributions are then combined, and the value of the $p_T$ cutoff separating the regions is varied to ensure that the total cross section is insensitive to this choice.

Figure~\ref{fig:ptcut} shows the dependence of the summed \texttt{MadGraph} and EPA contributions on the $p_T$ value separating the two regions, as well as the \texttt{MadGraph} contribution alone for the background process $\mu^+ \mu^+ \to \mu^+ \bar\nu_\mu W^+Z$. A plateau is observed between $1\ \mathrm{GeV}$ and $50\ \mathrm{GeV}$. We choose a $p_T$ cut of $0.6\ \mathrm{GeV}$ for the event generation, where the \texttt{MadGraph} contribution matches the correct value of the cross section.

\bibliographystyle{apsrev4-1}

\bibliography{Refs}

\begin{thebibliography}{98}%
\makeatletter
\providecommand \@ifxundefined [1]{%
 \@ifx{#1\undefined}
}%
\providecommand \@ifnum [1]{%
 \ifnum #1\expandafter \@firstoftwo
 \else \expandafter \@secondoftwo
 \fi
}%
\providecommand \@ifx [1]{%
 \ifx #1\expandafter \@firstoftwo
 \else \expandafter \@secondoftwo
 \fi
}%
\providecommand \natexlab [1]{#1}%
\providecommand \enquote  [1]{``#1''}%
\providecommand \bibnamefont  [1]{#1}%
\providecommand \bibfnamefont [1]{#1}%
\providecommand \citenamefont [1]{#1}%
\providecommand \href@noop [0]{\@secondoftwo}%
\providecommand \href [0]{\begingroup \@sanitize@url \@href}%
\providecommand \@href[1]{\@@startlink{#1}\@@href}%
\providecommand \@@href[1]{\endgroup#1\@@endlink}%
\providecommand \@sanitize@url [0]{\catcode `\\12\catcode `\$12\catcode `\&12\catcode `\#12\catcode `\^12\catcode `\_12\catcode `\%12\relax}%
\providecommand \@@startlink[1]{}%
\providecommand \@@endlink[0]{}%
\providecommand \url  [0]{\begingroup\@sanitize@url \@url }%
\providecommand \@url [1]{\endgroup\@href {#1}{\urlprefix }}%
\providecommand \urlprefix  [0]{URL }%
\providecommand \Eprint [0]{\href }%
\providecommand \doibase [0]{http://dx.doi.org/}%
\providecommand \selectlanguage [0]{\@gobble}%
\providecommand \bibinfo  [0]{\@secondoftwo}%
\providecommand \bibfield  [0]{\@secondoftwo}%
\providecommand \translation [1]{[#1]}%
\providecommand \BibitemOpen [0]{}%
\providecommand \bibitemStop [0]{}%
\providecommand \bibitemNoStop [0]{.\EOS\space}%
\providecommand \EOS [0]{\spacefactor3000\relax}%
\providecommand \BibitemShut  [1]{\csname bibitem#1\endcsname}%
\let\auto@bib@innerbib\@empty
\bibitem [{\citenamefont {Aad}\ \emph {et~al.}(2012)\citenamefont {Aad} \emph {et~al.}}]{ATLAS:2012yve}%
  \BibitemOpen
  \bibfield  {author} {\bibinfo {author} {\bibfnamefont {G.}~\bibnamefont {Aad}} \emph {et~al.} (\bibinfo {collaboration} {ATLAS}),\ }\href {\doibase 10.1016/j.physletb.2012.08.020} {\bibfield  {journal} {\bibinfo  {journal} {Phys. Lett. B}\ }\textbf {\bibinfo {volume} {716}},\ \bibinfo {pages} {1} (\bibinfo {year} {2012})},\ \Eprint {http://arxiv.org/abs/1207.7214} {arXiv:1207.7214 [hep-ex]} \BibitemShut {NoStop}%
\bibitem [{\citenamefont {Chatrchyan}\ \emph {et~al.}(2012)\citenamefont {Chatrchyan} \emph {et~al.}}]{CMS:2012qbp}%
  \BibitemOpen
  \bibfield  {author} {\bibinfo {author} {\bibfnamefont {S.}~\bibnamefont {Chatrchyan}} \emph {et~al.} (\bibinfo {collaboration} {CMS}),\ }\href {\doibase 10.1016/j.physletb.2012.08.021} {\bibfield  {journal} {\bibinfo  {journal} {Phys. Lett. B}\ }\textbf {\bibinfo {volume} {716}},\ \bibinfo {pages} {30} (\bibinfo {year} {2012})},\ \Eprint {http://arxiv.org/abs/1207.7235} {arXiv:1207.7235 [hep-ex]} \BibitemShut {NoStop}%
\bibitem [{\citenamefont {Benedikt}\ \emph {et~al.}(2025)\citenamefont {Benedikt} \emph {et~al.}}]{FCC:2025lpp}%
  \BibitemOpen
  \bibfield  {author} {\bibinfo {author} {\bibfnamefont {M.}~\bibnamefont {Benedikt}} \emph {et~al.} (\bibinfo {collaboration} {FCC}),\ }\href {\doibase 10.17181/CERN.9DKX.TDH9} {\  (\bibinfo {year} {2025}),\ 10.17181/CERN.9DKX.TDH9},\ \Eprint {http://arxiv.org/abs/2505.00272} {arXiv:2505.00272 [hep-ex]} \BibitemShut {NoStop}%
\bibitem [{\citenamefont {Behnke}\ \emph {et~al.}(2013)\citenamefont {Behnke} \emph {et~al.}}]{Behnke:2013xla}%
  \BibitemOpen
  \bibfield  {author} {\bibinfo {author} {\bibfnamefont {T.}~\bibnamefont {Behnke}} \emph {et~al.},\ }\href@noop {} {\  (\bibinfo {year} {2013})},\ \Eprint {http://arxiv.org/abs/1306.6327} {arXiv:1306.6327 [physics.acc-ph]} \BibitemShut {NoStop}%
\bibitem [{\citenamefont {Adolphsen}\ \emph {et~al.}(2013{\natexlab{a}})\citenamefont {Adolphsen} \emph {et~al.}}]{Adolphsen:2013jya}%
  \BibitemOpen
  \bibfield  {author} {\bibinfo {author} {\bibfnamefont {C.}~\bibnamefont {Adolphsen}} \emph {et~al.},\ }\href@noop {} {\  (\bibinfo {year} {2013}{\natexlab{a}})},\ \Eprint {http://arxiv.org/abs/1306.6353} {arXiv:1306.6353 [physics.acc-ph]} \BibitemShut {NoStop}%
\bibitem [{\citenamefont {Adolphsen}\ \emph {et~al.}(2013{\natexlab{b}})\citenamefont {Adolphsen} \emph {et~al.}}]{Adolphsen:2013kya}%
  \BibitemOpen
  \bibfield  {author} {\bibinfo {author} {\bibfnamefont {C.}~\bibnamefont {Adolphsen}} \emph {et~al.},\ }\href@noop {} {\  (\bibinfo {year} {2013}{\natexlab{b}})},\ \Eprint {http://arxiv.org/abs/1306.6328} {arXiv:1306.6328 [physics.acc-ph]} \BibitemShut {NoStop}%
\bibitem [{\citenamefont {Bambade}\ \emph {et~al.}(2019)\citenamefont {Bambade} \emph {et~al.}}]{Bambade:2019fyw}%
  \BibitemOpen
  \bibfield  {author} {\bibinfo {author} {\bibfnamefont {P.}~\bibnamefont {Bambade}} \emph {et~al.},\ }\href@noop {} {\  (\bibinfo {year} {2019})},\ \Eprint {http://arxiv.org/abs/1903.01629} {arXiv:1903.01629 [hep-ex]} \BibitemShut {NoStop}%
\bibitem [{\citenamefont {Zarnecki}(2020)}]{Zarnecki:2020ics}%
  \BibitemOpen
  \bibfield  {author} {\bibinfo {author} {\bibfnamefont {A.~F.}\ \bibnamefont {Zarnecki}} (\bibinfo {collaboration} {CLICdp, ILD concept group}),\ }\href {\doibase 10.22323/1.376.0037} {\bibfield  {journal} {\bibinfo  {journal} {PoS}\ }\textbf {\bibinfo {volume} {CORFU2019}},\ \bibinfo {pages} {037} (\bibinfo {year} {2020})},\ \Eprint {http://arxiv.org/abs/2004.14628} {arXiv:2004.14628 [hep-ph]} \BibitemShut {NoStop}%
\bibitem [{\citenamefont {Abramowicz}\ \emph {et~al.}(2025)\citenamefont {Abramowicz} \emph {et~al.}}]{LinearCollider:2025lya}%
  \BibitemOpen
  \bibfield  {author} {\bibinfo {author} {\bibfnamefont {H.}~\bibnamefont {Abramowicz}} \emph {et~al.} (\bibinfo {collaboration} {Linear Collider}),\ }\href@noop {} {\  (\bibinfo {year} {2025})},\ \Eprint {http://arxiv.org/abs/2503.24049} {arXiv:2503.24049 [hep-ex]} \BibitemShut {NoStop}%
\bibitem [{Aic(2018)}]{Aicheler:2018arh}%
  \BibitemOpen
  \href {\doibase 10.23731/CYRM-2018-004} {\ \textbf {\bibinfo {volume} {4/2018}} (\bibinfo {year} {2018}),\ 10.23731/CYRM-2018-004},\ \Eprint {http://arxiv.org/abs/1903.08655} {arXiv:1903.08655 [physics.acc-ph]} \BibitemShut {NoStop}%
\bibitem [{\citenamefont {Vernieri}\ \emph {et~al.}(2023)\citenamefont {Vernieri} \emph {et~al.}}]{Vernieri:2022fae}%
  \BibitemOpen
  \bibfield  {author} {\bibinfo {author} {\bibfnamefont {C.}~\bibnamefont {Vernieri}} \emph {et~al.},\ }\href {\doibase 10.1088/1748-0221/18/07/P07053} {\bibfield  {journal} {\bibinfo  {journal} {JINST}\ }\textbf {\bibinfo {volume} {18}},\ \bibinfo {pages} {P07053} (\bibinfo {year} {2023})},\ \Eprint {http://arxiv.org/abs/2203.07646} {arXiv:2203.07646 [hep-ex]} \BibitemShut {NoStop}%
\bibitem [{\citenamefont {Breidenbach}\ \emph {et~al.}(2023)\citenamefont {Breidenbach}, \citenamefont {Bullard}, \citenamefont {Nanni}, \citenamefont {Ntounis},\ and\ \citenamefont {Vernieri}}]{Breidenbach:2023nxd}%
  \BibitemOpen
  \bibfield  {author} {\bibinfo {author} {\bibfnamefont {M.}~\bibnamefont {Breidenbach}}, \bibinfo {author} {\bibfnamefont {B.}~\bibnamefont {Bullard}}, \bibinfo {author} {\bibfnamefont {E.~A.}\ \bibnamefont {Nanni}}, \bibinfo {author} {\bibfnamefont {D.}~\bibnamefont {Ntounis}}, \ and\ \bibinfo {author} {\bibfnamefont {C.}~\bibnamefont {Vernieri}},\ }\href {\doibase 10.1103/PRXEnergy.2.047001} {\bibfield  {journal} {\bibinfo  {journal} {PRX Energy}\ }\textbf {\bibinfo {volume} {2}},\ \bibinfo {pages} {047001} (\bibinfo {year} {2023})},\ \bibinfo {note} {[Erratum: PRX Energy 4, 029901 (2025)]},\ \Eprint {http://arxiv.org/abs/2307.04084} {arXiv:2307.04084 [hep-ex]} \BibitemShut {NoStop}%
\bibitem [{\citenamefont {Lindstr{\o}m}\ \emph {et~al.}(2023)\citenamefont {Lindstr{\o}m}, \citenamefont {D'Arcy},\ and\ \citenamefont {Foster}}]{Lindstrom:2023owp}%
  \BibitemOpen
  \bibfield  {author} {\bibinfo {author} {\bibfnamefont {C.~A.}\ \bibnamefont {Lindstr{\o}m}}, \bibinfo {author} {\bibfnamefont {R.}~\bibnamefont {D'Arcy}}, \ and\ \bibinfo {author} {\bibfnamefont {B.}~\bibnamefont {Foster}},\ }in\ \href@noop {} {\emph {\bibinfo {booktitle} {{6th European Advanced Accelerator Concepts workshop}}}}\ (\bibinfo {year} {2023})\ \Eprint {http://arxiv.org/abs/2312.04975} {arXiv:2312.04975 [physics.acc-ph]} \BibitemShut {NoStop}%
\bibitem [{\citenamefont {Foster}\ \emph {et~al.}(2023)\citenamefont {Foster}, \citenamefont {D'Arcy},\ and\ \citenamefont {Lindstrom}}]{Foster:2023bmq}%
  \BibitemOpen
  \bibfield  {author} {\bibinfo {author} {\bibfnamefont {B.}~\bibnamefont {Foster}}, \bibinfo {author} {\bibfnamefont {R.}~\bibnamefont {D'Arcy}}, \ and\ \bibinfo {author} {\bibfnamefont {C.~A.}\ \bibnamefont {Lindstrom}},\ }\href {\doibase 10.1088/1367-2630/acf395} {\bibfield  {journal} {\bibinfo  {journal} {New J. Phys.}\ }\textbf {\bibinfo {volume} {25}},\ \bibinfo {pages} {093037} (\bibinfo {year} {2023})},\ \Eprint {http://arxiv.org/abs/2303.10150} {arXiv:2303.10150 [physics.acc-ph]} \BibitemShut {NoStop}%
\bibitem [{\citenamefont {Dong}\ \emph {et~al.}(2018)\citenamefont {Dong} \emph {et~al.}}]{CEPCStudyGroup:2018ghi}%
  \BibitemOpen
  \bibfield  {author} {\bibinfo {author} {\bibfnamefont {M.}~\bibnamefont {Dong}} \emph {et~al.} (\bibinfo {collaboration} {CEPC Study Group}),\ }\href@noop {} {\  (\bibinfo {year} {2018})},\ \Eprint {http://arxiv.org/abs/1811.10545} {arXiv:1811.10545 [hep-ex]} \BibitemShut {NoStop}%
\bibitem [{\citenamefont {Anastopoulos}\ \emph {et~al.}(2025)\citenamefont {Anastopoulos} \emph {et~al.}}]{Anastopoulos:2025jyh}%
  \BibitemOpen
  \bibfield  {author} {\bibinfo {author} {\bibfnamefont {C.}~\bibnamefont {Anastopoulos}} \emph {et~al.},\ }\href@noop {} {\  (\bibinfo {year} {2025})},\ \Eprint {http://arxiv.org/abs/2504.00541} {arXiv:2504.00541 [physics.acc-ph]} \BibitemShut {NoStop}%
\bibitem [{\citenamefont {Accettura}\ \emph {et~al.}(2023)\citenamefont {Accettura} \emph {et~al.}}]{Accettura:2023ked}%
  \BibitemOpen
  \bibfield  {author} {\bibinfo {author} {\bibfnamefont {C.}~\bibnamefont {Accettura}} \emph {et~al.},\ }\href {\doibase 10.1140/epjc/s10052-023-11889-x} {\bibfield  {journal} {\bibinfo  {journal} {Eur. Phys. J. C}\ }\textbf {\bibinfo {volume} {83}},\ \bibinfo {pages} {864} (\bibinfo {year} {2023})},\ \bibinfo {note} {[Erratum: Eur.Phys.J.C 84, 36 (2024)]},\ \Eprint {http://arxiv.org/abs/2303.08533} {arXiv:2303.08533 [physics.acc-ph]} \BibitemShut {NoStop}%
\bibitem [{\citenamefont {Heusch}\ and\ \citenamefont {Cuypers}(1996)}]{Heusch:1995yw}%
  \BibitemOpen
  \bibfield  {author} {\bibinfo {author} {\bibfnamefont {C.~A.}\ \bibnamefont {Heusch}}\ and\ \bibinfo {author} {\bibfnamefont {F.}~\bibnamefont {Cuypers}},\ }\href {\doibase 10.1063/1.49345} {\bibfield  {journal} {\bibinfo  {journal} {AIP Conf. Proc.}\ }\textbf {\bibinfo {volume} {352}},\ \bibinfo {pages} {219} (\bibinfo {year} {1996})},\ \Eprint {http://arxiv.org/abs/hep-ph/9508230} {arXiv:hep-ph/9508230} \BibitemShut {NoStop}%
\bibitem [{\citenamefont {Antonelli}\ \emph {et~al.}(2016)\citenamefont {Antonelli}, \citenamefont {Boscolo}, \citenamefont {Di~Nardo},\ and\ \citenamefont {Raimondi}}]{Antonelli:2015nla}%
  \BibitemOpen
  \bibfield  {author} {\bibinfo {author} {\bibfnamefont {M.}~\bibnamefont {Antonelli}}, \bibinfo {author} {\bibfnamefont {M.}~\bibnamefont {Boscolo}}, \bibinfo {author} {\bibfnamefont {R.}~\bibnamefont {Di~Nardo}}, \ and\ \bibinfo {author} {\bibfnamefont {P.}~\bibnamefont {Raimondi}},\ }\href {\doibase 10.1016/j.nima.2015.10.097} {\bibfield  {journal} {\bibinfo  {journal} {Nucl. Instrum. Meth. A}\ }\textbf {\bibinfo {volume} {807}},\ \bibinfo {pages} {101} (\bibinfo {year} {2016})},\ \Eprint {http://arxiv.org/abs/1509.04454} {arXiv:1509.04454 [physics.acc-ph]} \BibitemShut {NoStop}%
\bibitem [{\citenamefont {Long}\ \emph {et~al.}(2021)\citenamefont {Long}, \citenamefont {Lucchesi}, \citenamefont {Palmer}, \citenamefont {Pastrone}, \citenamefont {Schulte},\ and\ \citenamefont {Shiltsev}}]{Long:2020wfp}%
  \BibitemOpen
  \bibfield  {author} {\bibinfo {author} {\bibfnamefont {K.}~\bibnamefont {Long}}, \bibinfo {author} {\bibfnamefont {D.}~\bibnamefont {Lucchesi}}, \bibinfo {author} {\bibfnamefont {M.}~\bibnamefont {Palmer}}, \bibinfo {author} {\bibfnamefont {N.}~\bibnamefont {Pastrone}}, \bibinfo {author} {\bibfnamefont {D.}~\bibnamefont {Schulte}}, \ and\ \bibinfo {author} {\bibfnamefont {V.}~\bibnamefont {Shiltsev}},\ }\href {\doibase 10.1038/s41567-020-01130-x} {\bibfield  {journal} {\bibinfo  {journal} {Nature Phys.}\ }\textbf {\bibinfo {volume} {17}},\ \bibinfo {pages} {289} (\bibinfo {year} {2021})},\ \Eprint {http://arxiv.org/abs/2007.15684} {arXiv:2007.15684 [physics.acc-ph]} \BibitemShut {NoStop}%
\bibitem [{\citenamefont {Delahaye}\ \emph {et~al.}(2019)\citenamefont {Delahaye}, \citenamefont {Diemoz}, \citenamefont {Long}, \citenamefont {Mansouli{\'e}}, \citenamefont {Pastrone}, \citenamefont {Rivkin}, \citenamefont {Schulte}, \citenamefont {Skrinsky},\ and\ \citenamefont {Wulzer}}]{Delahaye:2019omf}%
  \BibitemOpen
  \bibfield  {author} {\bibinfo {author} {\bibfnamefont {J.~P.}\ \bibnamefont {Delahaye}}, \bibinfo {author} {\bibfnamefont {M.}~\bibnamefont {Diemoz}}, \bibinfo {author} {\bibfnamefont {K.}~\bibnamefont {Long}}, \bibinfo {author} {\bibfnamefont {B.}~\bibnamefont {Mansouli{\'e}}}, \bibinfo {author} {\bibfnamefont {N.}~\bibnamefont {Pastrone}}, \bibinfo {author} {\bibfnamefont {L.}~\bibnamefont {Rivkin}}, \bibinfo {author} {\bibfnamefont {D.}~\bibnamefont {Schulte}}, \bibinfo {author} {\bibfnamefont {A.}~\bibnamefont {Skrinsky}}, \ and\ \bibinfo {author} {\bibfnamefont {A.}~\bibnamefont {Wulzer}},\ }\href@noop {} {\  (\bibinfo {year} {2019})},\ \Eprint {http://arxiv.org/abs/1901.06150} {arXiv:1901.06150 [physics.acc-ph]} \BibitemShut {NoStop}%
\bibitem [{\citenamefont {Stratakis}\ \emph {et~al.}(2022)\citenamefont {Stratakis} \emph {et~al.}}]{MuonCollider:2022nsa}%
  \BibitemOpen
  \bibfield  {author} {\bibinfo {author} {\bibfnamefont {D.}~\bibnamefont {Stratakis}} \emph {et~al.} (\bibinfo {collaboration} {Muon Collider}),\ }\href@noop {} {\  (\bibinfo {year} {2022})},\ \Eprint {http://arxiv.org/abs/2203.08033} {arXiv:2203.08033 [physics.acc-ph]} \BibitemShut {NoStop}%
\bibitem [{\citenamefont {Begel}\ \emph {et~al.}(2025)\citenamefont {Begel} \emph {et~al.}}]{Begel:2025ldu}%
  \BibitemOpen
  \bibfield  {author} {\bibinfo {author} {\bibfnamefont {M.}~\bibnamefont {Begel}} \emph {et~al.},\ }\href@noop {} {\  (\bibinfo {year} {2025})},\ \Eprint {http://arxiv.org/abs/2503.23695} {arXiv:2503.23695 [hep-ex]} \BibitemShut {NoStop}%
\bibitem [{\citenamefont {Kondo}\ \emph {et~al.}(2018)\citenamefont {Kondo} \emph {et~al.}}]{Kondo:2018rzx}%
  \BibitemOpen
  \bibfield  {author} {\bibinfo {author} {\bibfnamefont {Y.}~\bibnamefont {Kondo}} \emph {et~al.},\ }in\ \href {\doibase 10.18429/JACoW-IPAC2018-FRXGBF1} {\emph {\bibinfo {booktitle} {{9th International Particle Accelerator Conference}}}}\ (\bibinfo {year} {2018})\BibitemShut {NoStop}%
\bibitem [{\citenamefont {Aritome}\ \emph {et~al.}(2025)\citenamefont {Aritome} \emph {et~al.}}]{Aritome:2024rlu}%
  \BibitemOpen
  \bibfield  {author} {\bibinfo {author} {\bibfnamefont {S.}~\bibnamefont {Aritome}} \emph {et~al.},\ }\href {\doibase 10.1103/PhysRevLett.134.245001} {\bibfield  {journal} {\bibinfo  {journal} {Phys. Rev. Lett.}\ }\textbf {\bibinfo {volume} {134}},\ \bibinfo {pages} {245001} (\bibinfo {year} {2025})},\ \Eprint {http://arxiv.org/abs/2410.11367} {arXiv:2410.11367 [physics.acc-ph]} \BibitemShut {NoStop}%
\bibitem [{\citenamefont {Belanger}\ \emph {et~al.}(1996)\citenamefont {Belanger}, \citenamefont {Boudjema}, \citenamefont {London},\ and\ \citenamefont {Nadeau}}]{Belanger:1995nh}%
  \BibitemOpen
  \bibfield  {author} {\bibinfo {author} {\bibfnamefont {G.}~\bibnamefont {Belanger}}, \bibinfo {author} {\bibfnamefont {F.}~\bibnamefont {Boudjema}}, \bibinfo {author} {\bibfnamefont {D.}~\bibnamefont {London}}, \ and\ \bibinfo {author} {\bibfnamefont {H.}~\bibnamefont {Nadeau}},\ }\href {\doibase 10.1103/PhysRevD.53.6292} {\bibfield  {journal} {\bibinfo  {journal} {Phys. Rev. D}\ }\textbf {\bibinfo {volume} {53}},\ \bibinfo {pages} {6292} (\bibinfo {year} {1996})},\ \Eprint {http://arxiv.org/abs/hep-ph/9508317} {arXiv:hep-ph/9508317} \BibitemShut {NoStop}%
\bibitem [{\citenamefont {Gluza}\ and\ \citenamefont {Zralek}(1995)}]{Gluza:1995ix}%
  \BibitemOpen
  \bibfield  {author} {\bibinfo {author} {\bibfnamefont {J.}~\bibnamefont {Gluza}}\ and\ \bibinfo {author} {\bibfnamefont {M.}~\bibnamefont {Zralek}},\ }\href {\doibase 10.1016/0370-2693(95)01158-M} {\bibfield  {journal} {\bibinfo  {journal} {Phys. Lett. B}\ }\textbf {\bibinfo {volume} {362}},\ \bibinfo {pages} {148} (\bibinfo {year} {1995})},\ \Eprint {http://arxiv.org/abs/hep-ph/9507269} {arXiv:hep-ph/9507269} \BibitemShut {NoStop}%
\bibitem [{\citenamefont {Fridell}\ \emph {et~al.}(2023)\citenamefont {Fridell}, \citenamefont {Kitano},\ and\ \citenamefont {Takai}}]{Fridell:2023gjx}%
  \BibitemOpen
  \bibfield  {author} {\bibinfo {author} {\bibfnamefont {K.}~\bibnamefont {Fridell}}, \bibinfo {author} {\bibfnamefont {R.}~\bibnamefont {Kitano}}, \ and\ \bibinfo {author} {\bibfnamefont {R.}~\bibnamefont {Takai}},\ }\href {\doibase 10.1007/JHEP06(2023)086} {\bibfield  {journal} {\bibinfo  {journal} {JHEP}\ }\textbf {\bibinfo {volume} {06}},\ \bibinfo {pages} {086} (\bibinfo {year} {2023})},\ \Eprint {http://arxiv.org/abs/2304.14020} {arXiv:2304.14020 [hep-ph]} \BibitemShut {NoStop}%
\bibitem [{\citenamefont {Fukuda}\ \emph {et~al.}(2024)\citenamefont {Fukuda}, \citenamefont {Moroi}, \citenamefont {Niki},\ and\ \citenamefont {Wei}}]{Fukuda:2023yui}%
  \BibitemOpen
  \bibfield  {author} {\bibinfo {author} {\bibfnamefont {H.}~\bibnamefont {Fukuda}}, \bibinfo {author} {\bibfnamefont {T.}~\bibnamefont {Moroi}}, \bibinfo {author} {\bibfnamefont {A.}~\bibnamefont {Niki}}, \ and\ \bibinfo {author} {\bibfnamefont {S.-F.}\ \bibnamefont {Wei}},\ }\href {\doibase 10.1007/JHEP02(2024)214} {\bibfield  {journal} {\bibinfo  {journal} {JHEP}\ }\textbf {\bibinfo {volume} {02}},\ \bibinfo {pages} {214} (\bibinfo {year} {2024})},\ \Eprint {http://arxiv.org/abs/2310.07162} {arXiv:2310.07162 [hep-ph]} \BibitemShut {NoStop}%
\bibitem [{\citenamefont {Das}\ and\ \citenamefont {Orikasa}(2024)}]{Das:2024gfg}%
  \BibitemOpen
  \bibfield  {author} {\bibinfo {author} {\bibfnamefont {A.}~\bibnamefont {Das}}\ and\ \bibinfo {author} {\bibfnamefont {Y.}~\bibnamefont {Orikasa}},\ }\href {\doibase 10.1016/j.physletb.2024.138577} {\bibfield  {journal} {\bibinfo  {journal} {Phys. Lett. B}\ }\textbf {\bibinfo {volume} {851}},\ \bibinfo {pages} {138577} (\bibinfo {year} {2024})},\ \Eprint {http://arxiv.org/abs/2401.00696} {arXiv:2401.00696 [hep-ph]} \BibitemShut {NoStop}%
\bibitem [{\citenamefont {Kriewald}\ \emph {et~al.}(2025)\citenamefont {Kriewald}, \citenamefont {Pinsard},\ and\ \citenamefont {Teixeira}}]{Kriewald:2024cnt}%
  \BibitemOpen
  \bibfield  {author} {\bibinfo {author} {\bibfnamefont {J.}~\bibnamefont {Kriewald}}, \bibinfo {author} {\bibfnamefont {E.}~\bibnamefont {Pinsard}}, \ and\ \bibinfo {author} {\bibfnamefont {A.~M.}\ \bibnamefont {Teixeira}},\ }\href {\doibase 10.1007/JHEP02(2025)116} {\bibfield  {journal} {\bibinfo  {journal} {JHEP}\ }\textbf {\bibinfo {volume} {02}},\ \bibinfo {pages} {116} (\bibinfo {year} {2025})},\ \Eprint {http://arxiv.org/abs/2412.04331} {arXiv:2412.04331 [hep-ph]} \BibitemShut {NoStop}%
\bibitem [{\citenamefont {de~Lima}\ \emph {et~al.}(2025)\citenamefont {de~Lima}, \citenamefont {McKeen}, \citenamefont {Ng}, \citenamefont {Shamma},\ and\ \citenamefont {Tuckler}}]{deLima:2024ohf}%
  \BibitemOpen
  \bibfield  {author} {\bibinfo {author} {\bibfnamefont {C.~H.}\ \bibnamefont {de~Lima}}, \bibinfo {author} {\bibfnamefont {D.}~\bibnamefont {McKeen}}, \bibinfo {author} {\bibfnamefont {J.~N.}\ \bibnamefont {Ng}}, \bibinfo {author} {\bibfnamefont {M.}~\bibnamefont {Shamma}}, \ and\ \bibinfo {author} {\bibfnamefont {D.}~\bibnamefont {Tuckler}},\ }\href {\doibase 10.1103/PhysRevD.111.075002} {\bibfield  {journal} {\bibinfo  {journal} {Phys. Rev. D}\ }\textbf {\bibinfo {volume} {111}},\ \bibinfo {pages} {075002} (\bibinfo {year} {2025})},\ \Eprint {http://arxiv.org/abs/2411.15303} {arXiv:2411.15303 [hep-ph]} \BibitemShut {NoStop}%
\bibitem [{\citenamefont {Hamada}\ \emph {et~al.}(2022)\citenamefont {Hamada}, \citenamefont {Kitano}, \citenamefont {Matsudo}, \citenamefont {Takaura},\ and\ \citenamefont {Yoshida}}]{Hamada:2022mua}%
  \BibitemOpen
  \bibfield  {author} {\bibinfo {author} {\bibfnamefont {Y.}~\bibnamefont {Hamada}}, \bibinfo {author} {\bibfnamefont {R.}~\bibnamefont {Kitano}}, \bibinfo {author} {\bibfnamefont {R.}~\bibnamefont {Matsudo}}, \bibinfo {author} {\bibfnamefont {H.}~\bibnamefont {Takaura}}, \ and\ \bibinfo {author} {\bibfnamefont {M.}~\bibnamefont {Yoshida}},\ }\href {\doibase 10.1093/ptep/ptac059} {\bibfield  {journal} {\bibinfo  {journal} {PTEP}\ }\textbf {\bibinfo {volume} {2022}},\ \bibinfo {pages} {053B02} (\bibinfo {year} {2022})},\ \Eprint {http://arxiv.org/abs/2201.06664} {arXiv:2201.06664 [hep-ph]} \BibitemShut {NoStop}%
\bibitem [{\citenamefont {Das}\ \emph {et~al.}(2024)\citenamefont {Das}, \citenamefont {Li}, \citenamefont {Mandal}, \citenamefont {Nomura},\ and\ \citenamefont {Zhang}}]{Das:2024kyk}%
  \BibitemOpen
  \bibfield  {author} {\bibinfo {author} {\bibfnamefont {A.}~\bibnamefont {Das}}, \bibinfo {author} {\bibfnamefont {J.}~\bibnamefont {Li}}, \bibinfo {author} {\bibfnamefont {S.}~\bibnamefont {Mandal}}, \bibinfo {author} {\bibfnamefont {T.}~\bibnamefont {Nomura}}, \ and\ \bibinfo {author} {\bibfnamefont {R.}~\bibnamefont {Zhang}},\ }\href@noop {} {\  (\bibinfo {year} {2024})},\ \Eprint {http://arxiv.org/abs/2410.21956} {arXiv:2410.21956 [hep-ph]} \BibitemShut {NoStop}%
\bibitem [{\citenamefont {Zyla}\ \emph {et~al.}(2020)\citenamefont {Zyla} \emph {et~al.}}]{ParticleDataGroup:2020ssz}%
  \BibitemOpen
  \bibfield  {author} {\bibinfo {author} {\bibfnamefont {P.~A.}\ \bibnamefont {Zyla}} \emph {et~al.} (\bibinfo {collaboration} {Particle Data Group}),\ }\href {\doibase 10.1093/ptep/ptaa104} {\bibfield  {journal} {\bibinfo  {journal} {PTEP}\ }\textbf {\bibinfo {volume} {2020}},\ \bibinfo {pages} {083C01} (\bibinfo {year} {2020})}\BibitemShut {NoStop}%
\bibitem [{\citenamefont {Minkowski}(1977)}]{Minkowski:1977sc}%
  \BibitemOpen
  \bibfield  {author} {\bibinfo {author} {\bibfnamefont {P.}~\bibnamefont {Minkowski}},\ }\href {\doibase 10.1016/0370-2693(77)90435-X} {\bibfield  {journal} {\bibinfo  {journal} {Phys. Lett. B}\ }\textbf {\bibinfo {volume} {67}},\ \bibinfo {pages} {421} (\bibinfo {year} {1977})}\BibitemShut {NoStop}%
\bibitem [{\citenamefont {Yanagida}(1979)}]{Yanagida:1979as}%
  \BibitemOpen
  \bibfield  {author} {\bibinfo {author} {\bibfnamefont {T.}~\bibnamefont {Yanagida}},\ }\href@noop {} {\bibfield  {journal} {\bibinfo  {journal} {Conf. Proc. C}\ }\textbf {\bibinfo {volume} {7902131}},\ \bibinfo {pages} {95} (\bibinfo {year} {1979})}\BibitemShut {NoStop}%
\bibitem [{\citenamefont {Glashow}(1980)}]{Glashow:1979nm}%
  \BibitemOpen
  \bibfield  {author} {\bibinfo {author} {\bibfnamefont {S.~L.}\ \bibnamefont {Glashow}},\ }\href {\doibase 10.1007/978-1-4684-7197-7_15} {\bibfield  {journal} {\bibinfo  {journal} {NATO Sci. Ser. B}\ }\textbf {\bibinfo {volume} {61}},\ \bibinfo {pages} {687} (\bibinfo {year} {1980})}\BibitemShut {NoStop}%
\bibitem [{\citenamefont {Gell-Mann}\ \emph {et~al.}(1979)\citenamefont {Gell-Mann}, \citenamefont {Ramond},\ and\ \citenamefont {Slansky}}]{Gell-Mann:1979vob}%
  \BibitemOpen
  \bibfield  {author} {\bibinfo {author} {\bibfnamefont {M.}~\bibnamefont {Gell-Mann}}, \bibinfo {author} {\bibfnamefont {P.}~\bibnamefont {Ramond}}, \ and\ \bibinfo {author} {\bibfnamefont {R.}~\bibnamefont {Slansky}},\ }\href@noop {} {\bibfield  {journal} {\bibinfo  {journal} {Conf. Proc. C}\ }\textbf {\bibinfo {volume} {790927}},\ \bibinfo {pages} {315} (\bibinfo {year} {1979})},\ \Eprint {http://arxiv.org/abs/1306.4669} {arXiv:1306.4669 [hep-th]} \BibitemShut {NoStop}%
\bibitem [{\citenamefont {Mohapatra}\ and\ \citenamefont {Senjanovic}(1980)}]{Mohapatra:1979ia}%
  \BibitemOpen
  \bibfield  {author} {\bibinfo {author} {\bibfnamefont {R.~N.}\ \bibnamefont {Mohapatra}}\ and\ \bibinfo {author} {\bibfnamefont {G.}~\bibnamefont {Senjanovic}},\ }\href {\doibase 10.1103/PhysRevLett.44.912} {\bibfield  {journal} {\bibinfo  {journal} {Phys. Rev. Lett.}\ }\textbf {\bibinfo {volume} {44}},\ \bibinfo {pages} {912} (\bibinfo {year} {1980})}\BibitemShut {NoStop}%
\bibitem [{\citenamefont {Akhmedov}\ \emph {et~al.}(1996)\citenamefont {Akhmedov}, \citenamefont {Lindner}, \citenamefont {Schnapka},\ and\ \citenamefont {Valle}}]{Akhmedov:1995vm}%
  \BibitemOpen
  \bibfield  {author} {\bibinfo {author} {\bibfnamefont {E.~K.}\ \bibnamefont {Akhmedov}}, \bibinfo {author} {\bibfnamefont {M.}~\bibnamefont {Lindner}}, \bibinfo {author} {\bibfnamefont {E.}~\bibnamefont {Schnapka}}, \ and\ \bibinfo {author} {\bibfnamefont {J.~W.~F.}\ \bibnamefont {Valle}},\ }\href {\doibase 10.1103/PhysRevD.53.2752} {\bibfield  {journal} {\bibinfo  {journal} {Phys. Rev. D}\ }\textbf {\bibinfo {volume} {53}},\ \bibinfo {pages} {2752} (\bibinfo {year} {1996})},\ \Eprint {http://arxiv.org/abs/hep-ph/9509255} {arXiv:hep-ph/9509255} \BibitemShut {NoStop}%
\bibitem [{\citenamefont {Barr}(2004)}]{Barr:2003nn}%
  \BibitemOpen
  \bibfield  {author} {\bibinfo {author} {\bibfnamefont {S.~M.}\ \bibnamefont {Barr}},\ }\href {\doibase 10.1103/PhysRevLett.92.101601} {\bibfield  {journal} {\bibinfo  {journal} {Phys. Rev. Lett.}\ }\textbf {\bibinfo {volume} {92}},\ \bibinfo {pages} {101601} (\bibinfo {year} {2004})},\ \Eprint {http://arxiv.org/abs/hep-ph/0309152} {arXiv:hep-ph/0309152} \BibitemShut {NoStop}%
\bibitem [{\citenamefont {Malinsky}\ \emph {et~al.}(2005)\citenamefont {Malinsky}, \citenamefont {Romao},\ and\ \citenamefont {Valle}}]{Malinsky:2005bi}%
  \BibitemOpen
  \bibfield  {author} {\bibinfo {author} {\bibfnamefont {M.}~\bibnamefont {Malinsky}}, \bibinfo {author} {\bibfnamefont {J.~C.}\ \bibnamefont {Romao}}, \ and\ \bibinfo {author} {\bibfnamefont {J.~W.~F.}\ \bibnamefont {Valle}},\ }\href {\doibase 10.1103/PhysRevLett.95.161801} {\bibfield  {journal} {\bibinfo  {journal} {Phys. Rev. Lett.}\ }\textbf {\bibinfo {volume} {95}},\ \bibinfo {pages} {161801} (\bibinfo {year} {2005})},\ \Eprint {http://arxiv.org/abs/hep-ph/0506296} {arXiv:hep-ph/0506296} \BibitemShut {NoStop}%
\bibitem [{\citenamefont {Schechter}\ and\ \citenamefont {Valle}(1980)}]{Schechter:1980gr}%
  \BibitemOpen
  \bibfield  {author} {\bibinfo {author} {\bibfnamefont {J.}~\bibnamefont {Schechter}}\ and\ \bibinfo {author} {\bibfnamefont {J.~W.~F.}\ \bibnamefont {Valle}},\ }\href {\doibase 10.1103/PhysRevD.22.2227} {\bibfield  {journal} {\bibinfo  {journal} {Phys. Rev. D}\ }\textbf {\bibinfo {volume} {22}},\ \bibinfo {pages} {2227} (\bibinfo {year} {1980})}\BibitemShut {NoStop}%
\bibitem [{\citenamefont {Gronau}\ \emph {et~al.}(1984)\citenamefont {Gronau}, \citenamefont {Leung},\ and\ \citenamefont {Rosner}}]{Gronau:1984ct}%
  \BibitemOpen
  \bibfield  {author} {\bibinfo {author} {\bibfnamefont {M.}~\bibnamefont {Gronau}}, \bibinfo {author} {\bibfnamefont {C.~N.}\ \bibnamefont {Leung}}, \ and\ \bibinfo {author} {\bibfnamefont {J.~L.}\ \bibnamefont {Rosner}},\ }\href {\doibase 10.1103/PhysRevD.29.2539} {\bibfield  {journal} {\bibinfo  {journal} {Phys. Rev. D}\ }\textbf {\bibinfo {volume} {29}},\ \bibinfo {pages} {2539} (\bibinfo {year} {1984})}\BibitemShut {NoStop}%
\bibitem [{\citenamefont {Mohapatra}\ and\ \citenamefont {Valle}(1986)}]{Mohapatra:1986bd}%
  \BibitemOpen
  \bibfield  {author} {\bibinfo {author} {\bibfnamefont {R.~N.}\ \bibnamefont {Mohapatra}}\ and\ \bibinfo {author} {\bibfnamefont {J.~W.~F.}\ \bibnamefont {Valle}},\ }\href {\doibase 10.1103/PhysRevD.34.1642} {\bibfield  {journal} {\bibinfo  {journal} {Phys. Rev. D}\ }\textbf {\bibinfo {volume} {34}},\ \bibinfo {pages} {1642} (\bibinfo {year} {1986})}\BibitemShut {NoStop}%
\bibitem [{\citenamefont {Deppisch}\ \emph {et~al.}(2015)\citenamefont {Deppisch}, \citenamefont {Bhupal~Dev},\ and\ \citenamefont {Pilaftsis}}]{Deppisch:2015qwa}%
  \BibitemOpen
  \bibfield  {author} {\bibinfo {author} {\bibfnamefont {F.~F.}\ \bibnamefont {Deppisch}}, \bibinfo {author} {\bibfnamefont {P.~S.}\ \bibnamefont {Bhupal~Dev}}, \ and\ \bibinfo {author} {\bibfnamefont {A.}~\bibnamefont {Pilaftsis}},\ }\href {\doibase 10.1088/1367-2630/17/7/075019} {\bibfield  {journal} {\bibinfo  {journal} {New J. Phys.}\ }\textbf {\bibinfo {volume} {17}},\ \bibinfo {pages} {075019} (\bibinfo {year} {2015})},\ \Eprint {http://arxiv.org/abs/1502.06541} {arXiv:1502.06541 [hep-ph]} \BibitemShut {NoStop}%
\bibitem [{\citenamefont {Bolton}\ \emph {et~al.}(2020)\citenamefont {Bolton}, \citenamefont {Deppisch},\ and\ \citenamefont {Bhupal~Dev}}]{Bolton:2019pcu}%
  \BibitemOpen
  \bibfield  {author} {\bibinfo {author} {\bibfnamefont {P.~D.}\ \bibnamefont {Bolton}}, \bibinfo {author} {\bibfnamefont {F.~F.}\ \bibnamefont {Deppisch}}, \ and\ \bibinfo {author} {\bibfnamefont {P.~S.}\ \bibnamefont {Bhupal~Dev}},\ }\href {\doibase 10.1007/JHEP03(2020)170} {\bibfield  {journal} {\bibinfo  {journal} {JHEP}\ }\textbf {\bibinfo {volume} {03}},\ \bibinfo {pages} {170} (\bibinfo {year} {2020})},\ \Eprint {http://arxiv.org/abs/1912.03058} {arXiv:1912.03058 [hep-ph]} \BibitemShut {NoStop}%
\bibitem [{\citenamefont {Sirunyan}\ \emph {et~al.}(2018)\citenamefont {Sirunyan} \emph {et~al.}}]{CMS:2018iaf}%
  \BibitemOpen
  \bibfield  {author} {\bibinfo {author} {\bibfnamefont {A.~M.}\ \bibnamefont {Sirunyan}} \emph {et~al.} (\bibinfo {collaboration} {CMS}),\ }\href {\doibase 10.1103/PhysRevLett.120.221801} {\bibfield  {journal} {\bibinfo  {journal} {Phys. Rev. Lett.}\ }\textbf {\bibinfo {volume} {120}},\ \bibinfo {pages} {221801} (\bibinfo {year} {2018})},\ \Eprint {http://arxiv.org/abs/1802.02965} {arXiv:1802.02965 [hep-ex]} \BibitemShut {NoStop}%
\bibitem [{\citenamefont {Tumasyan}\ \emph {et~al.}(2023)\citenamefont {Tumasyan} \emph {et~al.}}]{CMS:2022hvh}%
  \BibitemOpen
  \bibfield  {author} {\bibinfo {author} {\bibfnamefont {A.}~\bibnamefont {Tumasyan}} \emph {et~al.} (\bibinfo {collaboration} {CMS}),\ }\href {\doibase 10.1103/PhysRevLett.131.011803} {\bibfield  {journal} {\bibinfo  {journal} {Phys. Rev. Lett.}\ }\textbf {\bibinfo {volume} {131}},\ \bibinfo {pages} {011803} (\bibinfo {year} {2023})},\ \Eprint {http://arxiv.org/abs/2206.08956} {arXiv:2206.08956 [hep-ex]} \BibitemShut {NoStop}%
\bibitem [{\citenamefont {Tumasyan}\ \emph {et~al.}(2022)\citenamefont {Tumasyan} \emph {et~al.}}]{CMS:2022fut}%
  \BibitemOpen
  \bibfield  {author} {\bibinfo {author} {\bibfnamefont {A.}~\bibnamefont {Tumasyan}} \emph {et~al.} (\bibinfo {collaboration} {CMS}),\ }\href {\doibase 10.1007/JHEP07(2022)081} {\bibfield  {journal} {\bibinfo  {journal} {JHEP}\ }\textbf {\bibinfo {volume} {07}},\ \bibinfo {pages} {081} (\bibinfo {year} {2022})},\ \Eprint {http://arxiv.org/abs/2201.05578} {arXiv:2201.05578 [hep-ex]} \BibitemShut {NoStop}%
\bibitem [{\citenamefont {Rizzo}(1982)}]{Rizzo:1982kn}%
  \BibitemOpen
  \bibfield  {author} {\bibinfo {author} {\bibfnamefont {T.~G.}\ \bibnamefont {Rizzo}},\ }\href {\doibase 10.1016/0370-2693(82)90027-2} {\bibfield  {journal} {\bibinfo  {journal} {Phys. Lett. B}\ }\textbf {\bibinfo {volume} {116}},\ \bibinfo {pages} {23} (\bibinfo {year} {1982})}\BibitemShut {NoStop}%
\bibitem [{\citenamefont {Tornow}(2014)}]{Tornow:2014vta}%
  \BibitemOpen
  \bibfield  {author} {\bibinfo {author} {\bibfnamefont {W.}~\bibnamefont {Tornow}} (\bibinfo {collaboration} {KamLAND-Zen}),\ }in\ \href@noop {} {\emph {\bibinfo {booktitle} {{34th International Symposium on Physics in Collision}}}}\ (\bibinfo {year} {2014})\ \Eprint {http://arxiv.org/abs/1412.0734} {arXiv:1412.0734 [nucl-ex]} \BibitemShut {NoStop}%
\bibitem [{\citenamefont {Anton}\ \emph {et~al.}(2019)\citenamefont {Anton} \emph {et~al.}}]{EXO-200:2019rkq}%
  \BibitemOpen
  \bibfield  {author} {\bibinfo {author} {\bibfnamefont {G.}~\bibnamefont {Anton}} \emph {et~al.} (\bibinfo {collaboration} {EXO-200}),\ }\href {\doibase 10.1103/PhysRevLett.123.161802} {\bibfield  {journal} {\bibinfo  {journal} {Phys. Rev. Lett.}\ }\textbf {\bibinfo {volume} {123}},\ \bibinfo {pages} {161802} (\bibinfo {year} {2019})},\ \Eprint {http://arxiv.org/abs/1906.02723} {arXiv:1906.02723 [hep-ex]} \BibitemShut {NoStop}%
\bibitem [{\citenamefont {Agostini}\ \emph {et~al.}(2020)\citenamefont {Agostini} \emph {et~al.}}]{GERDA:2020xhi}%
  \BibitemOpen
  \bibfield  {author} {\bibinfo {author} {\bibfnamefont {M.}~\bibnamefont {Agostini}} \emph {et~al.} (\bibinfo {collaboration} {GERDA}),\ }\href {\doibase 10.1103/PhysRevLett.125.252502} {\bibfield  {journal} {\bibinfo  {journal} {Phys. Rev. Lett.}\ }\textbf {\bibinfo {volume} {125}},\ \bibinfo {pages} {252502} (\bibinfo {year} {2020})},\ \Eprint {http://arxiv.org/abs/2009.06079} {arXiv:2009.06079 [nucl-ex]} \BibitemShut {NoStop}%
\bibitem [{\citenamefont {Adams}\ \emph {et~al.}(2022)\citenamefont {Adams} \emph {et~al.}}]{CUORE:2021mvw}%
  \BibitemOpen
  \bibfield  {author} {\bibinfo {author} {\bibfnamefont {D.~Q.}\ \bibnamefont {Adams}} \emph {et~al.} (\bibinfo {collaboration} {CUORE}),\ }\href {\doibase 10.1038/s41586-022-04497-4} {\bibfield  {journal} {\bibinfo  {journal} {Nature}\ }\textbf {\bibinfo {volume} {604}},\ \bibinfo {pages} {53} (\bibinfo {year} {2022})},\ \Eprint {http://arxiv.org/abs/2104.06906} {arXiv:2104.06906 [nucl-ex]} \BibitemShut {NoStop}%
\bibitem [{\citenamefont {Abe}\ \emph {et~al.}(2023)\citenamefont {Abe} \emph {et~al.}}]{KamLAND-Zen:2022tow}%
  \BibitemOpen
  \bibfield  {author} {\bibinfo {author} {\bibfnamefont {S.}~\bibnamefont {Abe}} \emph {et~al.} (\bibinfo {collaboration} {KamLAND-Zen}),\ }\href {\doibase 10.1103/PhysRevLett.130.051801} {\bibfield  {journal} {\bibinfo  {journal} {Phys. Rev. Lett.}\ }\textbf {\bibinfo {volume} {130}},\ \bibinfo {pages} {051801} (\bibinfo {year} {2023})},\ \Eprint {http://arxiv.org/abs/2203.02139} {arXiv:2203.02139 [hep-ex]} \BibitemShut {NoStop}%
\bibitem [{\citenamefont {Rodejohann}(2010)}]{Rodejohann:2010jh}%
  \BibitemOpen
  \bibfield  {author} {\bibinfo {author} {\bibfnamefont {W.}~\bibnamefont {Rodejohann}},\ }\href {\doibase 10.1103/PhysRevD.81.114001} {\bibfield  {journal} {\bibinfo  {journal} {Phys. Rev. D}\ }\textbf {\bibinfo {volume} {81}},\ \bibinfo {pages} {114001} (\bibinfo {year} {2010})},\ \Eprint {http://arxiv.org/abs/1005.2854} {arXiv:1005.2854 [hep-ph]} \BibitemShut {NoStop}%
\bibitem [{\citenamefont {Jiang}\ \emph {et~al.}(2024)\citenamefont {Jiang}, \citenamefont {Yang}, \citenamefont {Qian}, \citenamefont {Ban}, \citenamefont {Li}, \citenamefont {You},\ and\ \citenamefont {Li}}]{Jiang:2023mte}%
  \BibitemOpen
  \bibfield  {author} {\bibinfo {author} {\bibfnamefont {R.}~\bibnamefont {Jiang}}, \bibinfo {author} {\bibfnamefont {T.}~\bibnamefont {Yang}}, \bibinfo {author} {\bibfnamefont {S.}~\bibnamefont {Qian}}, \bibinfo {author} {\bibfnamefont {Y.}~\bibnamefont {Ban}}, \bibinfo {author} {\bibfnamefont {J.}~\bibnamefont {Li}}, \bibinfo {author} {\bibfnamefont {Z.}~\bibnamefont {You}}, \ and\ \bibinfo {author} {\bibfnamefont {Q.}~\bibnamefont {Li}},\ }\href {\doibase 10.1103/PhysRevD.109.035020} {\bibfield  {journal} {\bibinfo  {journal} {Phys. Rev. D}\ }\textbf {\bibinfo {volume} {109}},\ \bibinfo {pages} {035020} (\bibinfo {year} {2024})},\ \Eprint {http://arxiv.org/abs/2304.04483} {arXiv:2304.04483 [hep-ph]} \BibitemShut {NoStop}%
\bibitem [{\citenamefont {Yang}\ \emph {et~al.}(2024)\citenamefont {Yang}, \citenamefont {Chang},\ and\ \citenamefont {Feng}}]{Yang:2023ojm}%
  \BibitemOpen
  \bibfield  {author} {\bibinfo {author} {\bibfnamefont {J.-L.}\ \bibnamefont {Yang}}, \bibinfo {author} {\bibfnamefont {C.-H.}\ \bibnamefont {Chang}}, \ and\ \bibinfo {author} {\bibfnamefont {T.-F.}\ \bibnamefont {Feng}},\ }\href {\doibase 10.1088/1674-1137/ad17b0} {\bibfield  {journal} {\bibinfo  {journal} {Chin. Phys. C}\ }\textbf {\bibinfo {volume} {48}},\ \bibinfo {pages} {043101} (\bibinfo {year} {2024})},\ \Eprint {http://arxiv.org/abs/2302.13247} {arXiv:2302.13247 [hep-ph]} \BibitemShut {NoStop}%
\bibitem [{\citenamefont {Dehghani}\ \emph {et~al.}(2025)\citenamefont {Dehghani}, \citenamefont {Frank},\ and\ \citenamefont {Fuks}}]{Dehghani:2025xkd}%
  \BibitemOpen
  \bibfield  {author} {\bibinfo {author} {\bibfnamefont {P.}~\bibnamefont {Dehghani}}, \bibinfo {author} {\bibfnamefont {M.}~\bibnamefont {Frank}}, \ and\ \bibinfo {author} {\bibfnamefont {B.}~\bibnamefont {Fuks}},\ }\href {\doibase 10.1103/3sxk-glsw} {\bibfield  {journal} {\bibinfo  {journal} {Phys. Rev. D}\ }\textbf {\bibinfo {volume} {112}},\ \bibinfo {pages} {035020} (\bibinfo {year} {2025})},\ \Eprint {http://arxiv.org/abs/2506.06159} {arXiv:2506.06159 [hep-ph]} \BibitemShut {NoStop}%
\bibitem [{\citenamefont {Dev}\ \emph {et~al.}(2024)\citenamefont {Dev}, \citenamefont {Heeck},\ and\ \citenamefont {Thapa}}]{Dev:2023nha}%
  \BibitemOpen
  \bibfield  {author} {\bibinfo {author} {\bibfnamefont {P.~S.~B.}\ \bibnamefont {Dev}}, \bibinfo {author} {\bibfnamefont {J.}~\bibnamefont {Heeck}}, \ and\ \bibinfo {author} {\bibfnamefont {A.}~\bibnamefont {Thapa}},\ }\href {\doibase 10.1140/epjc/s10052-024-12496-0} {\bibfield  {journal} {\bibinfo  {journal} {Eur. Phys. J. C}\ }\textbf {\bibinfo {volume} {84}},\ \bibinfo {pages} {148} (\bibinfo {year} {2024})},\ \Eprint {http://arxiv.org/abs/2309.06463} {arXiv:2309.06463 [hep-ph]} \BibitemShut {NoStop}%
\bibitem [{\citenamefont {Han}\ \emph {et~al.}(2026)\citenamefont {Han}, \citenamefont {Ibarra}, \citenamefont {Roy},\ and\ \citenamefont {Sabov{\'a}}}]{Han:2026ufa}%
  \BibitemOpen
  \bibfield  {author} {\bibinfo {author} {\bibfnamefont {T.}~\bibnamefont {Han}}, \bibinfo {author} {\bibfnamefont {A.}~\bibnamefont {Ibarra}}, \bibinfo {author} {\bibfnamefont {S.}~\bibnamefont {Roy}}, \ and\ \bibinfo {author} {\bibfnamefont {M.}~\bibnamefont {Sabov{\'a}}},\ }\href@noop {} {\  (\bibinfo {year} {2026})},\ \Eprint {http://arxiv.org/abs/2608.10062} {arXiv:2608.10062 [hep-ph]} \BibitemShut {NoStop}%
\bibitem [{\citenamefont {Weinberg}(1979)}]{Weinberg:1979sa}%
  \BibitemOpen
  \bibfield  {author} {\bibinfo {author} {\bibfnamefont {S.}~\bibnamefont {Weinberg}},\ }\href {\doibase 10.1103/PhysRevLett.43.1566} {\bibfield  {journal} {\bibinfo  {journal} {Phys. Rev. Lett.}\ }\textbf {\bibinfo {volume} {43}},\ \bibinfo {pages} {1566} (\bibinfo {year} {1979})}\BibitemShut {NoStop}%
\bibitem [{\citenamefont {Foot}\ \emph {et~al.}(1989)\citenamefont {Foot}, \citenamefont {Lew}, \citenamefont {He},\ and\ \citenamefont {Joshi}}]{Foot:1988aq}%
  \BibitemOpen
  \bibfield  {author} {\bibinfo {author} {\bibfnamefont {R.}~\bibnamefont {Foot}}, \bibinfo {author} {\bibfnamefont {H.}~\bibnamefont {Lew}}, \bibinfo {author} {\bibfnamefont {X.~G.}\ \bibnamefont {He}}, \ and\ \bibinfo {author} {\bibfnamefont {G.~C.}\ \bibnamefont {Joshi}},\ }\href {\doibase 10.1007/BF01415558} {\bibfield  {journal} {\bibinfo  {journal} {Z. Phys. C}\ }\textbf {\bibinfo {volume} {44}},\ \bibinfo {pages} {441} (\bibinfo {year} {1989})}\BibitemShut {NoStop}%
\bibitem [{\citenamefont {Ma}(1998)}]{Ma:1998dn}%
  \BibitemOpen
  \bibfield  {author} {\bibinfo {author} {\bibfnamefont {E.}~\bibnamefont {Ma}},\ }\href {\doibase 10.1103/PhysRevLett.81.1171} {\bibfield  {journal} {\bibinfo  {journal} {Phys. Rev. Lett.}\ }\textbf {\bibinfo {volume} {81}},\ \bibinfo {pages} {1171} (\bibinfo {year} {1998})},\ \Eprint {http://arxiv.org/abs/hep-ph/9805219} {arXiv:hep-ph/9805219} \BibitemShut {NoStop}%
\bibitem [{\citenamefont {Asaka}\ and\ \citenamefont {Shaposhnikov}(2005)}]{Asaka:2005pn}%
  \BibitemOpen
  \bibfield  {author} {\bibinfo {author} {\bibfnamefont {T.}~\bibnamefont {Asaka}}\ and\ \bibinfo {author} {\bibfnamefont {M.}~\bibnamefont {Shaposhnikov}},\ }\href {\doibase 10.1016/j.physletb.2005.06.020} {\bibfield  {journal} {\bibinfo  {journal} {Phys. Lett. B}\ }\textbf {\bibinfo {volume} {620}},\ \bibinfo {pages} {17} (\bibinfo {year} {2005})},\ \Eprint {http://arxiv.org/abs/hep-ph/0505013} {arXiv:hep-ph/0505013} \BibitemShut {NoStop}%
\bibitem [{\citenamefont {Bondarenko}\ \emph {et~al.}(2018)\citenamefont {Bondarenko}, \citenamefont {Boyarsky}, \citenamefont {Gorbunov},\ and\ \citenamefont {Ruchayskiy}}]{Bondarenko:2018ptm}%
  \BibitemOpen
  \bibfield  {author} {\bibinfo {author} {\bibfnamefont {K.}~\bibnamefont {Bondarenko}}, \bibinfo {author} {\bibfnamefont {A.}~\bibnamefont {Boyarsky}}, \bibinfo {author} {\bibfnamefont {D.}~\bibnamefont {Gorbunov}}, \ and\ \bibinfo {author} {\bibfnamefont {O.}~\bibnamefont {Ruchayskiy}},\ }\href {\doibase 10.1007/JHEP11(2018)032} {\bibfield  {journal} {\bibinfo  {journal} {JHEP}\ }\textbf {\bibinfo {volume} {11}},\ \bibinfo {pages} {032} (\bibinfo {year} {2018})},\ \Eprint {http://arxiv.org/abs/1805.08567} {arXiv:1805.08567 [hep-ph]} \BibitemShut {NoStop}%
\bibitem [{\citenamefont {Abdullahi}\ \emph {et~al.}(2023)\citenamefont {Abdullahi} \emph {et~al.}}]{Abdullahi:2022jlv}%
  \BibitemOpen
  \bibfield  {author} {\bibinfo {author} {\bibfnamefont {A.~M.}\ \bibnamefont {Abdullahi}} \emph {et~al.},\ }\href {\doibase 10.1088/1361-6471/ac98f9} {\bibfield  {journal} {\bibinfo  {journal} {J. Phys. G}\ }\textbf {\bibinfo {volume} {50}},\ \bibinfo {pages} {020501} (\bibinfo {year} {2023})},\ \Eprint {http://arxiv.org/abs/2203.08039} {arXiv:2203.08039 [hep-ph]} \BibitemShut {NoStop}%
\bibitem [{\citenamefont {Fukugita}\ and\ \citenamefont {Yanagida}(1986)}]{Fukugita:1986hr}%
  \BibitemOpen
  \bibfield  {author} {\bibinfo {author} {\bibfnamefont {M.}~\bibnamefont {Fukugita}}\ and\ \bibinfo {author} {\bibfnamefont {T.}~\bibnamefont {Yanagida}},\ }\href {\doibase 10.1016/0370-2693(86)91126-3} {\bibfield  {journal} {\bibinfo  {journal} {Phys. Lett. B}\ }\textbf {\bibinfo {volume} {174}},\ \bibinfo {pages} {45} (\bibinfo {year} {1986})}\BibitemShut {NoStop}%
\bibitem [{\citenamefont {Fukugita}\ and\ \citenamefont {Yanagida}(2002)}]{Fukugita:2002hu}%
  \BibitemOpen
  \bibfield  {author} {\bibinfo {author} {\bibfnamefont {M.}~\bibnamefont {Fukugita}}\ and\ \bibinfo {author} {\bibfnamefont {T.}~\bibnamefont {Yanagida}},\ }\href {\doibase 10.1103/PhysRevLett.89.131602} {\bibfield  {journal} {\bibinfo  {journal} {Phys. Rev. Lett.}\ }\textbf {\bibinfo {volume} {89}},\ \bibinfo {pages} {131602} (\bibinfo {year} {2002})},\ \Eprint {http://arxiv.org/abs/hep-ph/0203194} {arXiv:hep-ph/0203194} \BibitemShut {NoStop}%
\bibitem [{\citenamefont {Boyarsky}\ \emph {et~al.}(2009)\citenamefont {Boyarsky}, \citenamefont {Ruchayskiy},\ and\ \citenamefont {Shaposhnikov}}]{Boyarsky:2009ix}%
  \BibitemOpen
  \bibfield  {author} {\bibinfo {author} {\bibfnamefont {A.}~\bibnamefont {Boyarsky}}, \bibinfo {author} {\bibfnamefont {O.}~\bibnamefont {Ruchayskiy}}, \ and\ \bibinfo {author} {\bibfnamefont {M.}~\bibnamefont {Shaposhnikov}},\ }\href {\doibase 10.1146/annurev.nucl.010909.083654} {\bibfield  {journal} {\bibinfo  {journal} {Ann. Rev. Nucl. Part. Sci.}\ }\textbf {\bibinfo {volume} {59}},\ \bibinfo {pages} {191} (\bibinfo {year} {2009})},\ \Eprint {http://arxiv.org/abs/0901.0011} {arXiv:0901.0011 [hep-ph]} \BibitemShut {NoStop}%
\bibitem [{\citenamefont {Davidson}\ \emph {et~al.}(2008)\citenamefont {Davidson}, \citenamefont {Nardi},\ and\ \citenamefont {Nir}}]{Davidson:2008bu}%
  \BibitemOpen
  \bibfield  {author} {\bibinfo {author} {\bibfnamefont {S.}~\bibnamefont {Davidson}}, \bibinfo {author} {\bibfnamefont {E.}~\bibnamefont {Nardi}}, \ and\ \bibinfo {author} {\bibfnamefont {Y.}~\bibnamefont {Nir}},\ }\href {\doibase 10.1016/j.physrep.2008.06.002} {\bibfield  {journal} {\bibinfo  {journal} {Phys. Rept.}\ }\textbf {\bibinfo {volume} {466}},\ \bibinfo {pages} {105} (\bibinfo {year} {2008})},\ \Eprint {http://arxiv.org/abs/0802.2962} {arXiv:0802.2962 [hep-ph]} \BibitemShut {NoStop}%
\bibitem [{\citenamefont {Granelli}\ \emph {et~al.}(2025)\citenamefont {Granelli}, \citenamefont {Meloni}, \citenamefont {Parriciatu}, \citenamefont {Penedo},\ and\ \citenamefont {Petcov}}]{Granelli:2025lds}%
  \BibitemOpen
  \bibfield  {author} {\bibinfo {author} {\bibfnamefont {A.}~\bibnamefont {Granelli}}, \bibinfo {author} {\bibfnamefont {D.}~\bibnamefont {Meloni}}, \bibinfo {author} {\bibfnamefont {M.}~\bibnamefont {Parriciatu}}, \bibinfo {author} {\bibfnamefont {J.~T.}\ \bibnamefont {Penedo}}, \ and\ \bibinfo {author} {\bibfnamefont {S.~T.}\ \bibnamefont {Petcov}},\ }\href@noop {} {\  (\bibinfo {year} {2025})},\ \Eprint {http://arxiv.org/abs/2505.21405} {arXiv:2505.21405 [hep-ph]} \BibitemShut {NoStop}%
\bibitem [{\citenamefont {del Aguila}\ \emph {et~al.}(2008)\citenamefont {del Aguila}, \citenamefont {de~Blas},\ and\ \citenamefont {Perez-Victoria}}]{delAguila:2008pw}%
  \BibitemOpen
  \bibfield  {author} {\bibinfo {author} {\bibfnamefont {F.}~\bibnamefont {del Aguila}}, \bibinfo {author} {\bibfnamefont {J.}~\bibnamefont {de~Blas}}, \ and\ \bibinfo {author} {\bibfnamefont {M.}~\bibnamefont {Perez-Victoria}},\ }\href {\doibase 10.1103/PhysRevD.78.013010} {\bibfield  {journal} {\bibinfo  {journal} {Phys. Rev. D}\ }\textbf {\bibinfo {volume} {78}},\ \bibinfo {pages} {013010} (\bibinfo {year} {2008})},\ \Eprint {http://arxiv.org/abs/0803.4008} {arXiv:0803.4008 [hep-ph]} \BibitemShut {NoStop}%
\bibitem [{\citenamefont {Akhmedov}\ \emph {et~al.}(2013)\citenamefont {Akhmedov}, \citenamefont {Kartavtsev}, \citenamefont {Lindner}, \citenamefont {Michaels},\ and\ \citenamefont {Smirnov}}]{Akhmedov:2013hec}%
  \BibitemOpen
  \bibfield  {author} {\bibinfo {author} {\bibfnamefont {E.}~\bibnamefont {Akhmedov}}, \bibinfo {author} {\bibfnamefont {A.}~\bibnamefont {Kartavtsev}}, \bibinfo {author} {\bibfnamefont {M.}~\bibnamefont {Lindner}}, \bibinfo {author} {\bibfnamefont {L.}~\bibnamefont {Michaels}}, \ and\ \bibinfo {author} {\bibfnamefont {J.}~\bibnamefont {Smirnov}},\ }\href {\doibase 10.1007/JHEP05(2013)081} {\bibfield  {journal} {\bibinfo  {journal} {JHEP}\ }\textbf {\bibinfo {volume} {05}},\ \bibinfo {pages} {081} (\bibinfo {year} {2013})},\ \Eprint {http://arxiv.org/abs/1302.1872} {arXiv:1302.1872 [hep-ph]} \BibitemShut {NoStop}%
\bibitem [{\citenamefont {de~Blas}(2013)}]{deBlas:2013gla}%
  \BibitemOpen
  \bibfield  {author} {\bibinfo {author} {\bibfnamefont {J.}~\bibnamefont {de~Blas}},\ }\href {\doibase 10.1051/epjconf/20136019008} {\bibfield  {journal} {\bibinfo  {journal} {EPJ Web Conf.}\ }\textbf {\bibinfo {volume} {60}},\ \bibinfo {pages} {19008} (\bibinfo {year} {2013})},\ \Eprint {http://arxiv.org/abs/1307.6173} {arXiv:1307.6173 [hep-ph]} \BibitemShut {NoStop}%
\bibitem [{\citenamefont {Basso}\ \emph {et~al.}(2014)\citenamefont {Basso}, \citenamefont {Fischer},\ and\ \citenamefont {van~der Bij}}]{Basso:2013jka}%
  \BibitemOpen
  \bibfield  {author} {\bibinfo {author} {\bibfnamefont {L.}~\bibnamefont {Basso}}, \bibinfo {author} {\bibfnamefont {O.}~\bibnamefont {Fischer}}, \ and\ \bibinfo {author} {\bibfnamefont {J.~J.}\ \bibnamefont {van~der Bij}},\ }\href {\doibase 10.1209/0295-5075/105/11001} {\bibfield  {journal} {\bibinfo  {journal} {EPL}\ }\textbf {\bibinfo {volume} {105}},\ \bibinfo {pages} {11001} (\bibinfo {year} {2014})},\ \Eprint {http://arxiv.org/abs/1310.2057} {arXiv:1310.2057 [hep-ph]} \BibitemShut {NoStop}%
\bibitem [{\citenamefont {Antusch}\ and\ \citenamefont {Fischer}(2014)}]{Antusch:2014woa}%
  \BibitemOpen
  \bibfield  {author} {\bibinfo {author} {\bibfnamefont {S.}~\bibnamefont {Antusch}}\ and\ \bibinfo {author} {\bibfnamefont {O.}~\bibnamefont {Fischer}},\ }\href {\doibase 10.1007/JHEP10(2014)094} {\bibfield  {journal} {\bibinfo  {journal} {JHEP}\ }\textbf {\bibinfo {volume} {10}},\ \bibinfo {pages} {094} (\bibinfo {year} {2014})},\ \Eprint {http://arxiv.org/abs/1407.6607} {arXiv:1407.6607 [hep-ph]} \BibitemShut {NoStop}%
\bibitem [{\citenamefont {Antusch}\ and\ \citenamefont {Fischer}(2015)}]{Antusch:2015mia}%
  \BibitemOpen
  \bibfield  {author} {\bibinfo {author} {\bibfnamefont {S.}~\bibnamefont {Antusch}}\ and\ \bibinfo {author} {\bibfnamefont {O.}~\bibnamefont {Fischer}},\ }\href {\doibase 10.1007/JHEP05(2015)053} {\bibfield  {journal} {\bibinfo  {journal} {JHEP}\ }\textbf {\bibinfo {volume} {05}},\ \bibinfo {pages} {053} (\bibinfo {year} {2015})},\ \Eprint {http://arxiv.org/abs/1502.05915} {arXiv:1502.05915 [hep-ph]} \BibitemShut {NoStop}%
\bibitem [{\citenamefont {Chrzaszcz}\ \emph {et~al.}(2020)\citenamefont {Chrzaszcz}, \citenamefont {Drewes}, \citenamefont {Gonzalo}, \citenamefont {Harz}, \citenamefont {Krishnamurthy},\ and\ \citenamefont {Weniger}}]{Chrzaszcz:2019inj}%
  \BibitemOpen
  \bibfield  {author} {\bibinfo {author} {\bibfnamefont {M.}~\bibnamefont {Chrzaszcz}}, \bibinfo {author} {\bibfnamefont {M.}~\bibnamefont {Drewes}}, \bibinfo {author} {\bibfnamefont {T.~E.}\ \bibnamefont {Gonzalo}}, \bibinfo {author} {\bibfnamefont {J.}~\bibnamefont {Harz}}, \bibinfo {author} {\bibfnamefont {S.}~\bibnamefont {Krishnamurthy}}, \ and\ \bibinfo {author} {\bibfnamefont {C.}~\bibnamefont {Weniger}},\ }\href {\doibase 10.1140/epjc/s10052-020-8073-9} {\bibfield  {journal} {\bibinfo  {journal} {Eur. Phys. J. C}\ }\textbf {\bibinfo {volume} {80}},\ \bibinfo {pages} {569} (\bibinfo {year} {2020})},\ \Eprint {http://arxiv.org/abs/1908.02302} {arXiv:1908.02302 [hep-ph]} \BibitemShut {NoStop}%
\bibitem [{\citenamefont {Bryman}\ \emph {et~al.}(2022)\citenamefont {Bryman}, \citenamefont {Cirigliano}, \citenamefont {Crivellin},\ and\ \citenamefont {Inguglia}}]{Bryman:2021teu}%
  \BibitemOpen
  \bibfield  {author} {\bibinfo {author} {\bibfnamefont {D.}~\bibnamefont {Bryman}}, \bibinfo {author} {\bibfnamefont {V.}~\bibnamefont {Cirigliano}}, \bibinfo {author} {\bibfnamefont {A.}~\bibnamefont {Crivellin}}, \ and\ \bibinfo {author} {\bibfnamefont {G.}~\bibnamefont {Inguglia}},\ }\href {\doibase 10.1146/annurev-nucl-110121-051223} {\bibfield  {journal} {\bibinfo  {journal} {Ann. Rev. Nucl. Part. Sci.}\ }\textbf {\bibinfo {volume} {72}},\ \bibinfo {pages} {69} (\bibinfo {year} {2022})},\ \Eprint {http://arxiv.org/abs/2111.05338} {arXiv:2111.05338 [hep-ph]} \BibitemShut {NoStop}%
\bibitem [{\citenamefont {Blennow}\ \emph {et~al.}(2023)\citenamefont {Blennow}, \citenamefont {Fern\'andez-Mart\'\i{}nez}, \citenamefont {Hern\'andez-Garc\'\i{}a}, \citenamefont {L\'opez-Pav\'on}, \citenamefont {Marcano},\ and\ \citenamefont {Naredo-Tuero}}]{Blennow:2023mqx}%
  \BibitemOpen
  \bibfield  {author} {\bibinfo {author} {\bibfnamefont {M.}~\bibnamefont {Blennow}}, \bibinfo {author} {\bibfnamefont {E.}~\bibnamefont {Fern\'andez-Mart\'\i{}nez}}, \bibinfo {author} {\bibfnamefont {J.}~\bibnamefont {Hern\'andez-Garc\'\i{}a}}, \bibinfo {author} {\bibfnamefont {J.}~\bibnamefont {L\'opez-Pav\'on}}, \bibinfo {author} {\bibfnamefont {X.}~\bibnamefont {Marcano}}, \ and\ \bibinfo {author} {\bibfnamefont {D.}~\bibnamefont {Naredo-Tuero}},\ }\href {\doibase 10.1007/JHEP08(2023)030} {\bibfield  {journal} {\bibinfo  {journal} {JHEP}\ }\textbf {\bibinfo {volume} {08}},\ \bibinfo {pages} {030} (\bibinfo {year} {2023})},\ \Eprint {http://arxiv.org/abs/2306.01040} {arXiv:2306.01040 [hep-ph]} \BibitemShut {NoStop}%
\bibitem [{\citenamefont {Gallardo}\ \emph {et~al.}(1996)\citenamefont {Gallardo} \emph {et~al.}}]{Gallardo:1996aa}%
  \BibitemOpen
  \bibfield  {author} {\bibinfo {author} {\bibfnamefont {J.~C.}\ \bibnamefont {Gallardo}} \emph {et~al.},\ }\href@noop {} {\bibfield  {journal} {\bibinfo  {journal} {eConf}\ }\textbf {\bibinfo {volume} {C960625}},\ \bibinfo {pages} {R4} (\bibinfo {year} {1996})}\BibitemShut {NoStop}%
\bibitem [{\citenamefont {Ankenbrandt}\ \emph {et~al.}(1999)\citenamefont {Ankenbrandt} \emph {et~al.}}]{Ankenbrandt:1999cta}%
  \BibitemOpen
  \bibfield  {author} {\bibinfo {author} {\bibfnamefont {C.~M.}\ \bibnamefont {Ankenbrandt}} \emph {et~al.},\ }\href {\doibase 10.1103/PhysRevSTAB.2.081001} {\bibfield  {journal} {\bibinfo  {journal} {Phys. Rev. ST Accel. Beams}\ }\textbf {\bibinfo {volume} {2}},\ \bibinfo {pages} {081001} (\bibinfo {year} {1999})},\ \Eprint {http://arxiv.org/abs/physics/9901022} {arXiv:physics/9901022} \BibitemShut {NoStop}%
\bibitem [{\citenamefont {London}\ \emph {et~al.}(1987)\citenamefont {London}, \citenamefont {Belanger},\ and\ \citenamefont {Ng}}]{London:1987nz}%
  \BibitemOpen
  \bibfield  {author} {\bibinfo {author} {\bibfnamefont {D.}~\bibnamefont {London}}, \bibinfo {author} {\bibfnamefont {G.}~\bibnamefont {Belanger}}, \ and\ \bibinfo {author} {\bibfnamefont {J.~N.}\ \bibnamefont {Ng}},\ }\href {\doibase 10.1016/0370-2693(87)90723-4} {\bibfield  {journal} {\bibinfo  {journal} {Phys. Lett. B}\ }\textbf {\bibinfo {volume} {188}},\ \bibinfo {pages} {155} (\bibinfo {year} {1987})}\BibitemShut {NoStop}%
\bibitem [{\citenamefont {Heusch}\ and\ \citenamefont {Minkowski}(1994)}]{Heusch:1993qu}%
  \BibitemOpen
  \bibfield  {author} {\bibinfo {author} {\bibfnamefont {C.~A.}\ \bibnamefont {Heusch}}\ and\ \bibinfo {author} {\bibfnamefont {P.}~\bibnamefont {Minkowski}},\ }\href {\doibase 10.1016/0550-3213(94)90576-2} {\bibfield  {journal} {\bibinfo  {journal} {Nucl. Phys. B}\ }\textbf {\bibinfo {volume} {416}},\ \bibinfo {pages} {3} (\bibinfo {year} {1994})}\BibitemShut {NoStop}%
\bibitem [{\citenamefont {Asaka}\ and\ \citenamefont {Tsuyuki}(2015)}]{Asaka:2015oia}%
  \BibitemOpen
  \bibfield  {author} {\bibinfo {author} {\bibfnamefont {T.}~\bibnamefont {Asaka}}\ and\ \bibinfo {author} {\bibfnamefont {T.}~\bibnamefont {Tsuyuki}},\ }\href {\doibase 10.1103/PhysRevD.92.094012} {\bibfield  {journal} {\bibinfo  {journal} {Phys. Rev. D}\ }\textbf {\bibinfo {volume} {92}},\ \bibinfo {pages} {094012} (\bibinfo {year} {2015})},\ \Eprint {http://arxiv.org/abs/1508.04937} {arXiv:1508.04937 [hep-ph]} \BibitemShut {NoStop}%
\bibitem [{\citenamefont {Wang}\ \emph {et~al.}(2017)\citenamefont {Wang}, \citenamefont {Xu},\ and\ \citenamefont {Zhang}}]{Wang:2016eln}%
  \BibitemOpen
  \bibfield  {author} {\bibinfo {author} {\bibfnamefont {K.}~\bibnamefont {Wang}}, \bibinfo {author} {\bibfnamefont {T.}~\bibnamefont {Xu}}, \ and\ \bibinfo {author} {\bibfnamefont {L.}~\bibnamefont {Zhang}},\ }\href {\doibase 10.1103/PhysRevD.95.075021} {\bibfield  {journal} {\bibinfo  {journal} {Phys. Rev. D}\ }\textbf {\bibinfo {volume} {95}},\ \bibinfo {pages} {075021} (\bibinfo {year} {2017})},\ \Eprint {http://arxiv.org/abs/1610.02618} {arXiv:1610.02618 [hep-ph]} \BibitemShut {NoStop}%
\bibitem [{\citenamefont {Alwall}\ \emph {et~al.}(2014)\citenamefont {Alwall}, \citenamefont {Frederix}, \citenamefont {Frixione}, \citenamefont {Hirschi}, \citenamefont {Maltoni}, \citenamefont {Mattelaer}, \citenamefont {Shao}, \citenamefont {Stelzer}, \citenamefont {Torrielli},\ and\ \citenamefont {Zaro}}]{Alwall:2014hca}%
  \BibitemOpen
  \bibfield  {author} {\bibinfo {author} {\bibfnamefont {J.}~\bibnamefont {Alwall}}, \bibinfo {author} {\bibfnamefont {R.}~\bibnamefont {Frederix}}, \bibinfo {author} {\bibfnamefont {S.}~\bibnamefont {Frixione}}, \bibinfo {author} {\bibfnamefont {V.}~\bibnamefont {Hirschi}}, \bibinfo {author} {\bibfnamefont {F.}~\bibnamefont {Maltoni}}, \bibinfo {author} {\bibfnamefont {O.}~\bibnamefont {Mattelaer}}, \bibinfo {author} {\bibfnamefont {H.~S.}\ \bibnamefont {Shao}}, \bibinfo {author} {\bibfnamefont {T.}~\bibnamefont {Stelzer}}, \bibinfo {author} {\bibfnamefont {P.}~\bibnamefont {Torrielli}}, \ and\ \bibinfo {author} {\bibfnamefont {M.}~\bibnamefont {Zaro}},\ }\href {\doibase 10.1007/JHEP07(2014)079} {\bibfield  {journal} {\bibinfo  {journal} {JHEP}\ }\textbf {\bibinfo {volume} {07}},\ \bibinfo {pages} {079} (\bibinfo {year} {2014})},\ \Eprint {http://arxiv.org/abs/1405.0301} {arXiv:1405.0301 [hep-ph]} \BibitemShut {NoStop}%
\bibitem [{\citenamefont {Alloul}\ \emph {et~al.}(2014)\citenamefont {Alloul}, \citenamefont {Christensen}, \citenamefont {Degrande}, \citenamefont {Duhr},\ and\ \citenamefont {Fuks}}]{Alloul:2013bka}%
  \BibitemOpen
  \bibfield  {author} {\bibinfo {author} {\bibfnamefont {A.}~\bibnamefont {Alloul}}, \bibinfo {author} {\bibfnamefont {N.~D.}\ \bibnamefont {Christensen}}, \bibinfo {author} {\bibfnamefont {C.}~\bibnamefont {Degrande}}, \bibinfo {author} {\bibfnamefont {C.}~\bibnamefont {Duhr}}, \ and\ \bibinfo {author} {\bibfnamefont {B.}~\bibnamefont {Fuks}},\ }\href {\doibase 10.1016/j.cpc.2014.04.012} {\bibfield  {journal} {\bibinfo  {journal} {Comput. Phys. Commun.}\ }\textbf {\bibinfo {volume} {185}},\ \bibinfo {pages} {2250} (\bibinfo {year} {2014})},\ \Eprint {http://arxiv.org/abs/1310.1921} {arXiv:1310.1921 [hep-ph]} \BibitemShut {NoStop}%
\bibitem [{\citenamefont {Sjostrand}\ \emph {et~al.}(2006)\citenamefont {Sjostrand}, \citenamefont {Mrenna},\ and\ \citenamefont {Skands}}]{Sjostrand:2006za}%
  \BibitemOpen
  \bibfield  {author} {\bibinfo {author} {\bibfnamefont {T.}~\bibnamefont {Sjostrand}}, \bibinfo {author} {\bibfnamefont {S.}~\bibnamefont {Mrenna}}, \ and\ \bibinfo {author} {\bibfnamefont {P.~Z.}\ \bibnamefont {Skands}},\ }\href {\doibase 10.1088/1126-6708/2006/05/026} {\bibfield  {journal} {\bibinfo  {journal} {JHEP}\ }\textbf {\bibinfo {volume} {05}},\ \bibinfo {pages} {026} (\bibinfo {year} {2006})},\ \Eprint {http://arxiv.org/abs/hep-ph/0603175} {arXiv:hep-ph/0603175} \BibitemShut {NoStop}%
\bibitem [{\citenamefont {Sjostrand}\ \emph {et~al.}(2008)\citenamefont {Sjostrand}, \citenamefont {Mrenna},\ and\ \citenamefont {Skands}}]{Sjostrand:2007gs}%
  \BibitemOpen
  \bibfield  {author} {\bibinfo {author} {\bibfnamefont {T.}~\bibnamefont {Sjostrand}}, \bibinfo {author} {\bibfnamefont {S.}~\bibnamefont {Mrenna}}, \ and\ \bibinfo {author} {\bibfnamefont {P.~Z.}\ \bibnamefont {Skands}},\ }\href {\doibase 10.1016/j.cpc.2008.01.036} {\bibfield  {journal} {\bibinfo  {journal} {Comput. Phys. Commun.}\ }\textbf {\bibinfo {volume} {178}},\ \bibinfo {pages} {852} (\bibinfo {year} {2008})},\ \Eprint {http://arxiv.org/abs/0710.3820} {arXiv:0710.3820 [hep-ph]} \BibitemShut {NoStop}%
\bibitem [{\citenamefont {de~Favereau}\ \emph {et~al.}(2014)\citenamefont {de~Favereau}, \citenamefont {Delaere}, \citenamefont {Demin}, \citenamefont {Giammanco}, \citenamefont {Lema\^\i{}tre}, \citenamefont {Mertens},\ and\ \citenamefont {Selvaggi}}]{deFavereau:2013fsa}%
  \BibitemOpen
  \bibfield  {author} {\bibinfo {author} {\bibfnamefont {J.}~\bibnamefont {de~Favereau}}, \bibinfo {author} {\bibfnamefont {C.}~\bibnamefont {Delaere}}, \bibinfo {author} {\bibfnamefont {P.}~\bibnamefont {Demin}}, \bibinfo {author} {\bibfnamefont {A.}~\bibnamefont {Giammanco}}, \bibinfo {author} {\bibfnamefont {V.}~\bibnamefont {Lema\^\i{}tre}}, \bibinfo {author} {\bibfnamefont {A.}~\bibnamefont {Mertens}}, \ and\ \bibinfo {author} {\bibfnamefont {M.}~\bibnamefont {Selvaggi}} (\bibinfo {collaboration} {DELPHES 3}),\ }\href {\doibase 10.1007/JHEP02(2014)057} {\bibfield  {journal} {\bibinfo  {journal} {JHEP}\ }\textbf {\bibinfo {volume} {02}},\ \bibinfo {pages} {057} (\bibinfo {year} {2014})},\ \Eprint {http://arxiv.org/abs/1307.6346} {arXiv:1307.6346 [hep-ex]} \BibitemShut {NoStop}%
\bibitem [{\citenamefont {Hamada}\ \emph {et~al.}(2024)\citenamefont {Hamada}, \citenamefont {Kitano}, \citenamefont {Matsudo}, \citenamefont {Okawa}, \citenamefont {Takai}, \citenamefont {Takaura},\ and\ \citenamefont {Treuer}}]{Hamada:2024ojj}%
  \BibitemOpen
  \bibfield  {author} {\bibinfo {author} {\bibfnamefont {Y.}~\bibnamefont {Hamada}}, \bibinfo {author} {\bibfnamefont {R.}~\bibnamefont {Kitano}}, \bibinfo {author} {\bibfnamefont {R.}~\bibnamefont {Matsudo}}, \bibinfo {author} {\bibfnamefont {S.}~\bibnamefont {Okawa}}, \bibinfo {author} {\bibfnamefont {R.}~\bibnamefont {Takai}}, \bibinfo {author} {\bibfnamefont {H.}~\bibnamefont {Takaura}}, \ and\ \bibinfo {author} {\bibfnamefont {L.}~\bibnamefont {Treuer}},\ }\href {\doibase 10.1103/PhysRevD.110.113011} {\bibfield  {journal} {\bibinfo  {journal} {Phys. Rev. D}\ }\textbf {\bibinfo {volume} {110}},\ \bibinfo {pages} {113011} (\bibinfo {year} {2024})},\ \Eprint {http://arxiv.org/abs/2408.01068} {arXiv:2408.01068 [hep-ph]} \BibitemShut {NoStop}%
\bibitem [{\citenamefont {Harigaya}\ \emph {et~al.}(2025)\citenamefont {Harigaya}, \citenamefont {Kitano},\ and\ \citenamefont {Takai}}]{Harigaya:2025zru}%
  \BibitemOpen
  \bibfield  {author} {\bibinfo {author} {\bibfnamefont {K.}~\bibnamefont {Harigaya}}, \bibinfo {author} {\bibfnamefont {R.}~\bibnamefont {Kitano}}, \ and\ \bibinfo {author} {\bibfnamefont {R.}~\bibnamefont {Takai}},\ }\href@noop {} {\  (\bibinfo {year} {2025})},\ \Eprint {http://arxiv.org/abs/2509.24680} {arXiv:2509.24680 [hep-ph]} \BibitemShut {NoStop}%
\bibitem [{\citenamefont {von Weizsacker}(1934)}]{vonWeizsacker:1934nji}%
  \BibitemOpen
  \bibfield  {author} {\bibinfo {author} {\bibfnamefont {C.~F.}\ \bibnamefont {von Weizsacker}},\ }\href {\doibase 10.1007/BF01333110} {\bibfield  {journal} {\bibinfo  {journal} {Z. Phys.}\ }\textbf {\bibinfo {volume} {88}},\ \bibinfo {pages} {612} (\bibinfo {year} {1934})}\BibitemShut {NoStop}%
\bibitem [{\citenamefont {Williams}(1934)}]{Williams:1934ad}%
  \BibitemOpen
  \bibfield  {author} {\bibinfo {author} {\bibfnamefont {E.~J.}\ \bibnamefont {Williams}},\ }\href {\doibase 10.1103/PhysRev.45.729} {\bibfield  {journal} {\bibinfo  {journal} {Phys. Rev.}\ }\textbf {\bibinfo {volume} {45}},\ \bibinfo {pages} {729} (\bibinfo {year} {1934})}\BibitemShut {NoStop}%
\end{thebibliography}%
\end{document}